\documentclass[12pt]{article}

\usepackage{graphicx}
\usepackage{dcolumn}
\usepackage{bm}
\usepackage[a4paper,top=3.5cm,bottom=3.2cm,left=2.3cm,right=2.3cm,bindingoffset=5mm]{geometry}
\usepackage{amsfonts}
\usepackage{latexsym}
\usepackage{amsmath}
\usepackage{amssymb}
\usepackage[english]{babel}
\usepackage[latin1]{inputenc}
\usepackage[T1]{fontenc}
\usepackage{slashed}
\usepackage{float}
\usepackage{subfigure}
\usepackage{color}

    \newcommand{\be}{\begin{equation}}
    \newcommand{\ee}{\end{equation}}
    \newcommand{\ba}{\begin{eqnarray}}
    \newcommand{\ea}{\end{eqnarray}}
    \newcommand{\eite}{\end{itemize}}
    \newcommand{\bite}{\begin{itemize}}

\newcommand{\betabeta}{\mbox{$(\beta \beta)_{0 \nu}  $}}
\def\ltap{\ \raisebox{-.4ex}{\rlap{$\sim$}} \raisebox{.4ex}{$<$}\ }

\newcommand{\meff}{\mbox{$\left|  < \!  m \!  > \right| \ $}}
\newcommand{\mefff}{\mbox{$ < \! m \! > $}}

\newcommand{\eq}{\begin{eqnarray}}
\newcommand{\en}{\end{eqnarray}}

\newcommand{\nn}{\nonumber}

\begin{document}

\hfill{{\small Ref. SISSA 37/2012/EP}}
%
%
%
\vspace{1.0cm}
\begin{center}
{\bf{\large Multiple CP Non-conserving
Mechanisms of $\betabeta$-Decay and Nuclei with
Largely Different Nuclear Matrix Elements}}

\vspace{0.4cm}
A. Meroni$\mbox{}^{b,c)}$,
S. T. Petcov$\mbox{}^{b,c,d)}$
\footnote{Also at: Institute of Nuclear Research and
Nuclear Energy, Bulgarian Academy of Sciences, 1784 Sofia, Bulgaria}
and
F. \v Simkovic$\mbox{}^{e,f)}$

\vspace{0.1cm}
$\mbox{}^{b)}${\em  SISSA, Via Bonomea 265, 34136 Trieste, Italy.\\}

\vspace{0.1cm}
$\mbox{}^{c)}${\em  INFN, Sezione di Trieste, 34126 Trieste, Italy.\\}

\vspace{0.1cm}
$\mbox{}^{d)}${\em Kavli IPMU (WPI), The University of Tokyo, Kashiwa,
Japan.\\
}

\vspace{0.1cm}
$\mbox{}^{e)}${\em Department of Nuclear Physics and Biophysics, Comenius
University, Mlynska dolina F1, SK-842 15 Bratislava, Slovakia\\
}

\vspace{0.1cm}
$\mbox{}^{f)}${\em Bogoliubov Laboratory of Theoretical Physics, JINR,
141980 Dubna, Moscow region, Russia.\\
}

\end{center}

\begin{abstract}
We investigate the possibility to discriminate between
different pairs of CP non-conserving mechanisms inducing the
neutrinoless double beta $\betabeta$-decay by using
data on $\betabeta$-decay half-lives of nuclei
with largely different nuclear matrix elements (NMEs).
The mechanisms studied are: light Majorana neutrino exchange,
heavy left-handed (LH) and heavy right-handed (RH)
Majorana neutrino exchanges, lepton charge non-conserving
couplings in SUSY theories with $R$-parity breaking giving rise
to the ``dominant gluino exchange'' and the ``squark-neutrino''
mechanisms. The nuclei considered are
$^{76}$Ge, $^{82}$Se, $^{100}$Mo, $^{130}$Te
and $^{136}$Xe. Four sets of nuclear matrix
elements (NMEs) of the decays of these
five nuclei, derived within the
Self-consistent Renormalized
Quasiparticle Random Phase Approximation
(SRQRPA), were employed in our analysis.
While for each of the five single
mechanisms discussed, the NMEs for $^{76}$Ge, $^{82}$Se,
$^{100}$Mo and $^{130}$Te differ relatively little,
the relative difference between the NMEs of any two
nuclei not exceeding 10\%, the NMEs for $^{136}Xe$
differ significantly from those of
$^{76}$Ge, $^{82}Se$, $^{100}$Mo and $^{130}$Te,
being by a factor $\sim (1.3 - 2.5)$ smaller.
This allows, in principle, to draw conclusions about
the pair of non-interfering (interfering)
mechanisms possibly inducing the $\betabeta$-decay
from data on the half-lives of $^{136}Xe$ and
of at least one (two) more isotope(s) which can
be, e.g., any of the four,
$^{76}Ge$, $^{82}Se$, $^{100}Mo$ and $^{130}Te$.
Depending on the sets of mechanisms
considered, the conclusion can be independent of, 
or can depend on, the NMEs used in the analysis.
The implications of the EXO lower bound on the half-life of
$^{136}Xe$ for the problem studied are also exploited.

\end{abstract}


%
\section{Introduction}
%
%
\indent If neutrinoless double beta $\betabeta$-decay
will be observed, it will be of fundamental
importance to determine the mechanism which
induces the decay. In \cite{FMPSV0311}
we have considered the possibility of several
different mechanisms contributing to the neutrinoless
double beta $\betabeta$-decay amplitude in the
general case of CP nonconservation
\footnote{The case of two CP conserving mechanisms
generating the $\betabeta$-decay was considered
in \cite{FSV10MM}.}.
The mechanisms discussed are the ``standard''
light Majorana neutrino exchange,
exchange of heavy Majorana neutrinos
coupled to (V-A) currents, exchange of
heavy right-handed (RH) Majorana neutrinos
coupled to (V+A) currents,
lepton charge non-conserving couplings in SUSY theories
with $R$-parity breaking
\footnote{For a more detailed description of
these mechanisms and references to the articles
where they were originally proposed
see \cite{FMPSV0311} and, e.g.,
\cite{BiPet87,WRode0511}.}.
Of the latter we have
concentrated on the so-called
``dominant gluino exchange''
and ``squark-neutrino'' mechanisms.
Each of these mechanisms is characterized
by a specific fundamental lepton
number violating (LNV) parameter  $\eta_{\kappa}$,
the index $\kappa$ labeling the mechanism.
The parameter $\eta_{\kappa}$ will be complex,
in general,  if the mechanism $\kappa$
does not conserve the CP symmetry.
In \cite{Deppisch:2006hb,Gehman:2007qg,Fogli:2009py}
the authors analised the possibility to 
identify the mechanisms of  $\betabeta$-decay 
using data on the half-lives of several isotopes. 
The indicated (and other) specific mechanisms were considered, 
assuming  that only one of these mechanisms is triggering the decay 
\footnote{In \cite{HPas1999,HPas2001} the authors derived
a general effective $\betabeta$-decay Lagrangian   
without specifying the mechanisms 
generating the different terms in the Lagrangian.
Assuming  that only one of the terms in the Lagrangian 
is operative in $\betabeta$-decay 
and using the existing lower limit 
on the half-life of  $^{76}$Ge, 
in \cite{HPas2001} constraints on the 
effective CP conserving couplings constant multiplying 
the different terms in the Lagrangian were obtained.
}.
In \cite{FMPSV0311} we have investigated in detail
the cases of two ``non-interfering'' \cite{HPR83}
and two ``interfering'' mechanisms
generating the $\betabeta$-decay
\footnote{Two mechanisms contributing to the $\betabeta$-decay
amplitude can be non-interfering or interfering,
depending on whether their interference term present in the
$\betabeta$-decay rate is suppressed (and negligible) or not.
In the case of the five mechanisms considered in \cite{FMPSV0311}
and listed above, all pairs of mechanisms, which include
the exchange of the heavy RH Majorana neutrinos coupled
to (V+A) currents as one of the mechanisms,
can be shown to be of the non-interfering
type, while those that do not involve the
exchange of heavy RH Majorana neutrinos belong
to the interfering class.},
using as input hypothetical $\betabeta$-decay
half-lives of the three isotopes $^{76}$Ge,
$^{100}$Mo and $^{130}$Te.
Four sets of nuclear matrix
elements (NMEs) of the decays of these
three nuclei, derived within the
Self-consistent Renormalized
Quasiparticle Random Phase Approximation
(SRQRPA) \cite{srpa,ocup09}, were utilized:
they were calculated in \cite{FMPSV0311}
with two different nucleon-nucleon (NN) potentials
(CD-Bonn and Argonne), ``large'' size single-particle spaces 
and for two values of the axial coupling 
constant $g_A=1.25;~1.0$.

  If the $\betabeta$-decay is induced by
two non-interfering mechanisms,
which for concreteness were considered
in \cite{FMPSV0311} to be
\cite{HPR83} the light left-handed (LH) 
and the heavy RH Majorana neutrino exchanges, 
one can determine the squares of the 
absolute values of the two LNV parameters, 
characterizing these mechanisms,
$|\eta_\nu|^2$ and $|\eta_R|^2$, 
from data on the half-lives of two nuclear isotopes. 
This was done in \cite{FMPSV0311} 
using as input all three possible 
pairs of half-lives of  $^{76}$Ge, 
$^{100}$Mo and $^{130}$Te, chosen from specific intervals 
and satisfying the existing experimental constraints. 
It was found that if the half-life
of one of  the three nuclei is measured,
the requirement that
$|\eta_\nu|^2 \geq 0$ and $|\eta_R|^2 \geq 0$ 
(``positivity condition'')
constrains  the other two half-lives
(and the $\betabeta$-decay  half-life of any other
$\betabeta$-decaying isotope for that matter)
to lie in specific  intervals,
determined by the measured half-life and
the relevant NMEs and phase-space factors.
This  feature is common to all cases
of  two non-interfering mechanisms
generating the $\betabeta$-decay.
The indicated  specific
half-life intervals for the various isotopes
were shown to be stable with respect to the
change of the NMEs  (within the sets of
NMEs employed) used to derive them.
The intervals depend, in general,
on the type of the two
non-interfering mechanisms assumed
to cause the $\betabeta$-decay.
However, these differences in the cases
of all possible pairs of non-interfering
mechanisms considered were found to be extremely  small.
Using the indicated difference
to get information about the specific pair of
non-interfering  mechanisms
possibly operative in $\betabeta$-decay
requires, in the cases studied in \cite{FMPSV0311}, an
extremely high precision in the measurement of the
$\betabeta$-decay half-lives of the
isotopes considered ($^{76}$Ge, $^{100}$Mo and $^{130}$Te),
as well as an exceedingly small uncertainties
in the knowledge of the $\betabeta$-decay NMEs
of these isotopes. The levels of precision required
seem impossible to achieve in the foreseeable future.
One of the consequences of this
results is that if it will be possible
to rule out one pair of the considered in \cite{FMPSV0311}
non-interfering mechanisms
as the cause of  $\betabeta$-decay,
most likely one will  be able to rule out
all of them.

 The dependence of the physical solutions for
$|\eta_\nu|^2$ and  $|\eta_R|^2$
obtained on the NMEs used, was also studied
in \cite{FMPSV0311}.
It was found that the solutions
can exhibit a significant, or a relatively small,
variation with the NMEs employed, depending on
the hypothetical values of the half-lives of the two
isotopes utilized as input for obtaining the solutions.
This conclusion is valid for all other pairs
of non-interfering mechanisms considered
in \cite{FMPSV0311}.
In the case when two
interfering mechanisms are responsible
for the $\betabeta$-decay, the squares of the
absolute values of the two relevant parameters and the
interference term parameter, which involves
the cosine of an unknown relative phase $\alpha$
of the two fundamental parameters,
can be uniquely determined,
in principle, from data on the half-lives
of three nuclei. We have analyzed
in \cite{FMPSV0311} in detail the case
of light Majorana neutrino exchange
and gluino exchange.
In this case the parameters which are determined
from data on the half-lives
are $|\eta_\nu|^2$, $|\eta_{\lambda'}|^2$,
$\eta_{\lambda'}$ being that of the gluino exchange,
and $z = 2\cos\alpha\,|\eta_\nu||\eta_{\lambda'}|$.
The physical solutions for these parameters
have to satisfy the  conditions
$|\eta_\nu|^2 \geq 0$, $|\eta_{\lambda'}|^2\geq 0$
and $-\,2|\eta_\nu||\eta_{\lambda'}|\leq z \leq
2|\eta_\nu||\eta_{\lambda'}|$.
The latter condition implies that
given the half-lives of two isotopes, $T_1$ and $T_2$,
the half-life of any third isotope $T_3$
is constrained to lie is a specific interval,
if the mechanisms considered are indeed
generating the $\betabeta$-decay.
If further the half-life of one isotope $T_1$  is known,
for the interference to be constructive (destructive),
the half-lives of any other pair of
isotopes $T_2$ and $T_3$, should belong to
specific intervals. These intervals depend
on whether the interference
between the two contributions in the
$\betabeta$-decay rate
is constructive or destructive.
We have derived in \cite{FMPSV0311}
in analytic form  the general  conditions
for i) constructive interference ($z > 0$),
ii) destructive interference ($z< 0$),
iii)   $|\eta_\nu|^2 = 0$, $|\eta_{\lambda'}|^2 \neq 0$,
iv)  $|\eta_\nu|^2 \neq 0$,  $|\eta_{\lambda'}|^2 = 0$ and
v) $z = 0$,  $|\eta_\nu|^2 \neq 0$, $|\eta_{\lambda'}|^2 \neq 0$.
We have found that,  given $T_1$, a constructive
interference is possible only if $T_2$ lies in a relatively
narrow interval and $T_3$ has a value in extremely
narrow intervals. Numerically the intervals
for $T_2$ and $T_3$ are very similar to the intervals
one obtains in the case of two non-interfering
mechanisms (within the set considered
in \cite{FMPSV0311}).
The intervals  of values of $T_2$ and $T_3$
corresponding to destructive interference
are very different from those corresponding
to the cases of constructive interference and
of the  two non-interfering  $\betabeta$-decay
mechanisms we have considered.
Within the set of $\betabeta$-decay mechanisms studied by us,
this difference can allow to discriminate
experimentally between the  possibilities of the
$\betabeta$-decay being triggered by two `` destructively interfering''
mechanisms or  by two ``constructively  interfering''
or by two non-interfering  mechanisms.

 The ``degeneracy'' of the predictions of
the pairs of non-interfering
mechanisms of $\betabeta$-decay
for the interval of
values of the half-life of a second nucleus,
given the half-life of a different one
from the three considered,
$^{76}$Ge, $^{100}$Mo and $^{130}$Te,
is a direct consequence of a specific
property of the NMEs of the three nuclei.
Namely, for each of the five single
mechanisms discussed (the light LH,
heavy LH and heavy RH Majorana
neutrino exchanges, the gluino exchange and the
``squark-neutrino'' mechanism),
the NMEs for the three nuclei differ
relatively little, the relative difference between
the NMEs of any two nuclei not exceeding 10\%~
\footnote{The general implications of this ``degeneracy''
of the NMEs of $^{76}$Ge, $^{100}$Mo and $^{130}$Te
for testing the mechanisms under discussion in the case of
CP invariance were investigated in \cite{ELisietalMM11}.
}:
$|M_{i,\kappa} - M_{j,\kappa}|/(0.5 |M_{i,\kappa} + M_{j,\kappa}|)
\ltap 0.1$, $i\neq j=1,2,3\equiv ^{76}Ge,^{100}Mo,^{130}Te$.
This feature of the NMEs of $^{76}$Ge, $^{100}$Mo and $^{130}$Te
makes it impossible to discriminate
experimentally not only between the
four different pairs of non-interfering mechanisms,
considered in \cite{FMPSV0311},
using data on the half-lives of
$^{76}$Ge, $^{100}$Mo and $^{130}$Te,
but also between any of these four pairs
and pairs of interfering mechanisms
when the interference is constructive.
The indicated ``degeneracy'' of the predictions of
different pairs of mechanisms possibly active in
$\betabeta$-decay can be lifted if one uses
as input the half-lives of nuclei having largely different NMEs.
One example of such a nucleus is $^{136}Xe$,
whose NMEs for the five mechanisms studied in \cite{FMPSV0311},
as we are going to show,
differ significantly from those of
$^{76}$Ge, $^{82}Se$, $^{100}$Mo and $^{130}$Te.

   In the present article we investigate
the potential of combining data on the half-lives of
$^{136}Xe$ and of one or more of the
four nuclei $^{76}$Ge, $^{82}Se$, $^{100}$Mo and $^{130}$Te,
for discriminating between different pairs of non-interfering or
interfering mechanisms of $\betabeta$-decay.
We consider the same five basic mechanisms
used in the study performed in \cite{FMPSV0311},
namely, the ``standard''
light Majorana neutrino exchange,
exchange of heavy Majorana neutrinos
coupled to (V-A) currents, exchange of
heavy right-handed (RH) Majorana neutrinos
coupled to (V+A) currents, dominant gluino exchange
and the squark-neutrino  mechanism.
The last two are related to
lepton charge non-conserving couplings in SUSY theories
with $R$-parity breaking.

The paper is organized as follows. In Section 2 we give
a brief overview  on the possible mechanisms that can induce the 
$\betabeta$-decay considered in this work. In Section 3 
we analyze the case of two mechanisms active in the decay
in the non-interfering and in the interfering regime, 
combining the recent experimental results reported by EXO, 
while the Section 4 contains our conclusions. 

%
\section{Mechanisms of $\betabeta$-Decay Considered and
Nuclear Matrix Elements Employed}
\label{LNVmechanisms}
%
%

 The mechanisms of $\betabeta$-decay we are going
to consider in the present article are the same five
mechanisms considered in \cite{FMPSV0311}:
i) the light Majorana neutrino exchange,
characterized by the dimensionless LNV parameter
$\eta_{\nu} = \mefff/m_e$, $\mefff$ and and $m_e$ being
the $\betabeta$-decay effective Majorana mass
(see, e.g., \cite{BiPet87,WRode0511}) and
the electron mass, respectively;
ii) exchange of heavy ``left-handed'' (LH)
Majorana neutrinos coupled to (V-A) currents,
characterized by the LNV parameter $\eta_L$;
iii) exchange of heavy ``right-handed''
(RH) Majorana neutrinos coupled to (V+A) currents,
characterized by the LNV parameter $\eta_R$.
We consider also two  possible mechanisms 
in SUSY theories with R-parity 
 \begin{figure}[h!]
  \begin{center}
 {\includegraphics[width=10cm]{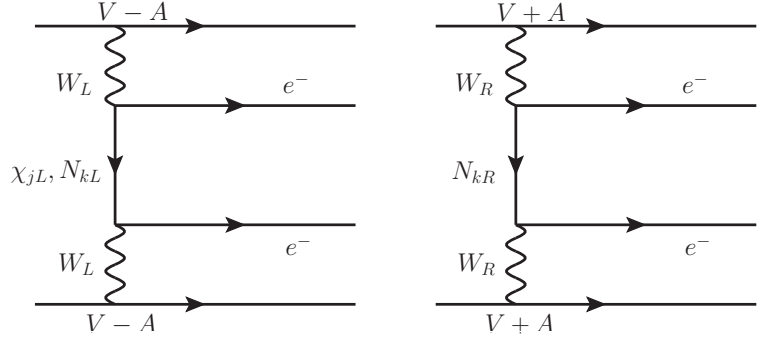}}
  \end{center}
\caption{\label{Feyn1} Feynman diagrams for the
$\betabeta$-decay, generated by the
light and heavy LH Majorana neutrino exchange
(left panel) and the heavy RH 
Majorana neutrino exchange (right panel).}
\end{figure}
 \begin{figure}[h!]
  \begin{center}
\subfigure
{\includegraphics[width=4.2cm]{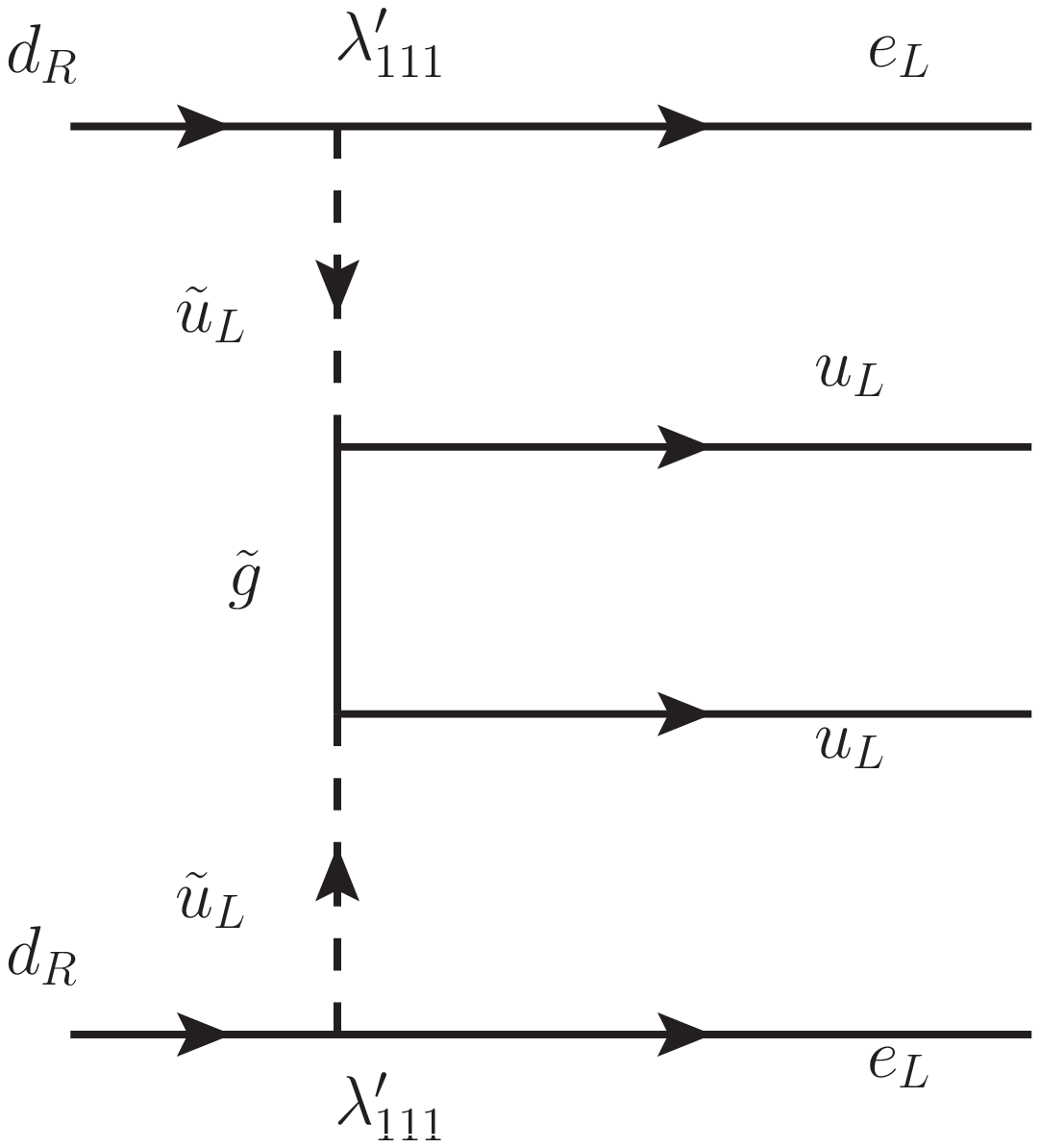}}
\hspace{20pt}
\subfigure
   {\includegraphics[width=4.7cm]{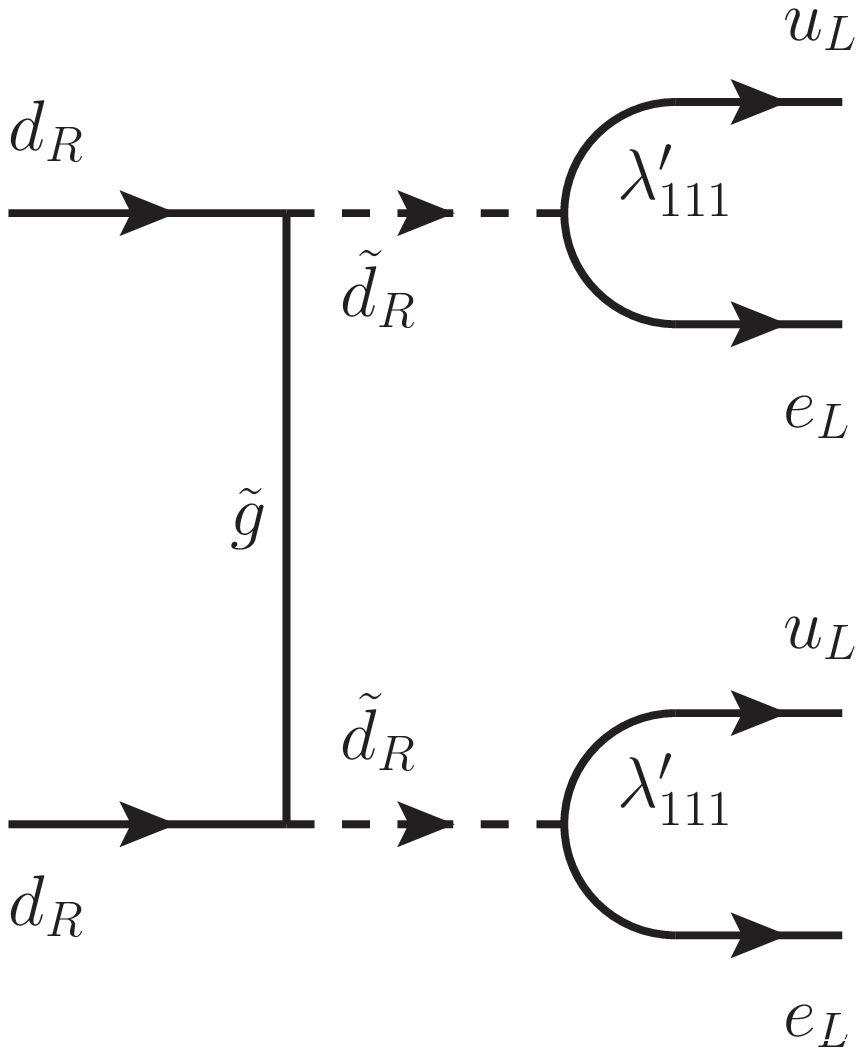}}
\hspace{20pt}
\subfigure
   {\includegraphics[width=4.2cm]{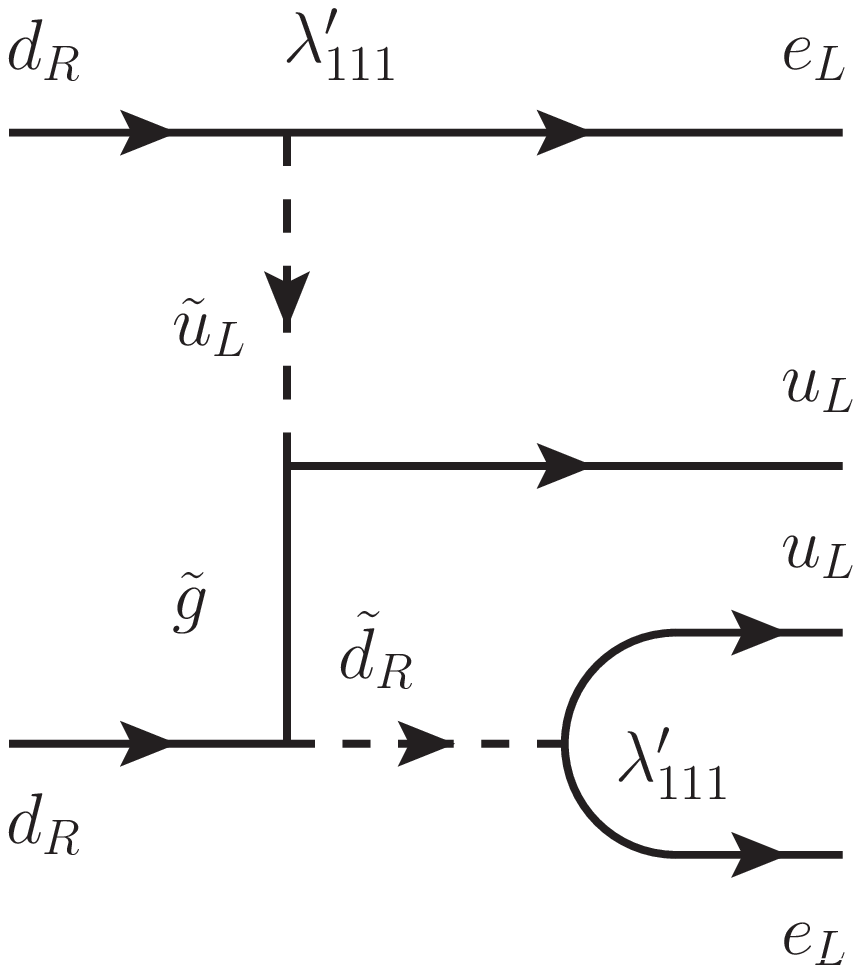}}
   \end{center}
     \caption{\label{Feyn2} Feynman diagrams
for $\betabeta$-decay due to the gluino exchange mechanism.
}
\end{figure}
%
non-conservation: iv) the ``gluino exchange'' mechanisms, characterized
by the LNV parameter $\eta_{\lambda'}$, and
v) the ``squark-neutrino'' mechanism, characterized
by the parameter $\eta_{\nu-q}$.
All five LNV parameters we have introduced above
 $\eta_{\nu}$,  $\eta_L$, $\eta_R$,
$\eta_{\lambda'}$ and  $\eta_{\nu-q}$ are effective.
The dependence of a given LNV effective parameter
on the fundamental parameters of the theory
in which the corresponding mechanism is possible/arises,
is discussed in detail in  \cite{FMPSV0311}~\footnote{An 
extensive list of references to the
original articles is also given in \cite{FMPSV0311}.}
and we are not going to repeat this discussion here.
 \begin{figure}[h!]
  \begin{center}
\subfigure
{\includegraphics[width=6cm]{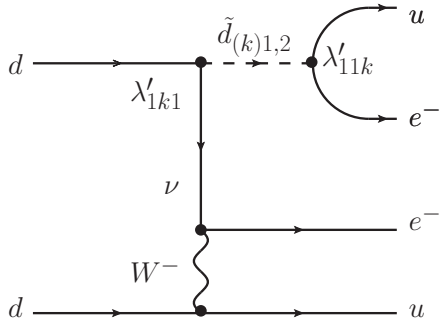}}
   \end{center}
     \caption{\label{Feyn3} Feynman diagrams for $\betabeta$-decay 
due to the squark-neutrino mechanism at the quark-level 
\cite{squark}.
}
\end{figure}
%
Instead we will give just a graphical representation
of each of the five mechanisms in terms of the
corresponding leading order Feynman diagrams.
For the light Majorana neutrino, heavy LH Majorana
neutrino ($N_{kL}$) and heavy RH Majorana
neutrino ($N_{kR}$) exchange mechanisms
they are given in Fig. \ref{Feyn1};
for the gluino exchange and squark-neutrino
mechanisms we show them in Figs. \ref{Feyn2}
and  \ref{Feyn3}.

 In the present analysis we employ four sets of nuclear matrix
elements (NMEs) of the decays of the 
five nuclei of interest,  $^{76}$Ge, $^{82}Se$, $^{100}$Mo, $^{130}$Te
and  $^{136}$Xe, derived within the
Self-consistent Renormalized Quasiparticle
Random Phase Approximation (SRQRPA) \cite{srpa,ocup09}.
The SRQRPA takes into account the Pauli
exclusion principle and conserves the mean
particle number in correlated ground state.

  For each of the five nuclei, two choices of single-particle
basis are considered. The intermediate size model space has 12
levels (oscillator shells N=2-4) for ${^{76}Ge}$ and ${^{82}Se}$, 16
levels (oscillator shells N=2-4 plus the f+h orbits from N=5) for
${^{100}Mo}$  and 18 levels (oscillator shells N=3,4 plus f+h+p
orbits from N=5) for  ${^{130}Te}$ and  ${^{136}Xe}$. 
The large size single particle space contains 21 levels 
(oscillator shells N=0-5)  for
${^{76}Ge}$, ${^{82}Se}$ and ${^{100}Mo}$, and 23 levels for
${^{130}Te}$ and ${^{136}Xe}$ (N=1-5 and $i$ orbits from N=6). 
In comparison with previous studies \cite{Rodin}, 
we omitted the small space model which is not sufficient 
to describe realistically the tensor part of 
the $\betabeta$-decay nuclear matrix elements.

  The single particle energies were obtained by using
a  Coulomb--corrected Woods--Saxon potential.
Two-body G-matrix elements we derived from
the Argonne and the Charge Dependent Bonn (CD-Bonn)
one-boson exchange potential within the Brueckner theory.
The schematic pairing interactions have been
adjusted to fit the empirical pairing gaps
\cite{cheo93}. The particle-particle and
particle-hole channels of the G-matrix
interaction of the nuclear Hamiltonian $H$ are renormalized by
introducing the parameters $g_{pp}$ and $g_{ph}$, respectively.
The calculations have been carried out for $g_{ph} = 1.0$.
The particle-particle strength parameter $g_{pp}$
of the SRQRPA is fixed by the data on the two-neutrino double
beta decays \cite{Rodin,anatomy}.
In the calculation of the
$\betabeta$-decay NMEs,
the two-nucleon short-range correlations
derived from same potential as residual interactions,
namely from the Argonne or CD-Bonn potentials, were considered
\cite{src}.

The calculated NMEs ${M'}^{0\nu}_\nu$, ${M'}^{0\nu}_N$,
${M'}^{0\nu}_{\lambda'}$ and ${M'}^{0\nu}_{\tilde q}$ are listed
in Table \ref{table.1}. For  $^{76}$Ge, $^{82}Se$, $^{100}$Mo
and $^{130}$Te they are taken from ref. \cite{FMPSV0311},
while for  $^{136}$Xe the results are new \cite{VES12}.
We note that these NMEs are significantly smaller (by a factor
1.3 - 2.5) when compared with those for  $^{76}$Ge, $^{82}Se$ 
and $^{100}$Mo. The reduction of the $0\nu\beta\beta$-decay 
NMEs of the $^{136}$Xe is explained  by the closed neutron 
shell for this nucleus. A sharper  Fermi surface leads to 
a reduction of this transition. This effect is clearly 
seen also in the case of ${M'}^{0\nu}_\nu$ of double 
magic nucleus $^{48}$Ca \cite{VES12}.

By glancing the Table \ref{table.1} we see that a significant source
of uncertainty is the value of the axial-vector coupling
constant $g_A$ and especially in the case of matrix elements
${M'}^{0\nu}_{\lambda'}$ and ${M'}^{0\nu}_{\tilde q}$. Further, 
the NMEs associated with heavy neutrino exchange are sensitive 
also to the choice of the NN interaction,
the CD-Bonn or Argonne potential. These types of realistic
NN interaction differ mostly by the description of the
short-range interactions. Although in Table \ref{table.1}
we present results for NMEs of the nuclei of interest,
calculated using both medium and large size single particle
spaces within the SRQRPA method, in the numerical examples
we are going to present further we will use
the NMEs for a given nucleus, with the large size
single particle space in both cases of  Argonne and CD-Bonn
\begin{table*}[htb]
  \begin{center}
\caption{\label{table.1} Nuclear matrix elements ${M'}^{0\nu}_\nu$
(light neutrino mass mechanism), ${M'}^{0\nu}_N$ (heavy neutrino
mass mechanism), ${M'}^{0\nu}_{\lambda'}$ (trilinear R-parity
breaking SUSY mechanism)  and ${M'}^{0\nu}_{\tilde q}$ (squark
mixing mechanism) for the $0\nu\beta\beta$-decays of $^{76}Ge$,
$^{100}Se$, $^{100}Mo$, $^{130}Te$  and $^{136}Xe$  within the
Selfconsistent Renormalized Quasiparticle Random Phase Approximation
(SRQRPA). $G^{0\nu}(E_0,Z)$ is the phase-space factor. 
We notice that all NMEs given in Table \ref{table.1} are real and
positive.  The nuclear radius is R = 1.1 fm A$^{1/3}$.
}\vspace{5pt}
\renewcommand{\tabcolsep}{1.1mm}
 \renewcommand{\arraystretch}{1}
{\footnotesize
\begin{tabular}{lcccccccccccccc}
\hline\hline
 Nuclear & $G^{0\nu}(E_0,Z)$ & & & \multicolumn{2}{c}{$|{M'}^{0\nu}_\nu|$} &  & \multicolumn{2}{c}{$|{M'}^{0\nu}_N|$} & &
\multicolumn{2}{c}{$|{M'}^{0\nu}_{\lambda'}|$} & & \multicolumn{2}{c}{$|{M'}^{0\nu}_{\tilde q}|$} \\
\cline{5-6} \cline{8-9} \cline{11-12} \cline{14-15}
 transition & [$y^{-1}$] & &  & \multicolumn{2}{c}{$g_A =$}  & & \multicolumn{2}{c}{$g_A =$}  & & \multicolumn{2}{c}{$g_A =$}
& & \multicolumn{2}{c}{$g_A =$} \\
   & & NN pot. & m.s.  & 1.0 & 1.25  & & 1.0 & 1.25 & & 1.0 & 1.25 & & 1.0 & 1.25 \\\hline
   & & &
    & & & & & & & & & & &    \\
$^{76}Ge\rightarrow {^{76}Se}$ & $7.98~10^{-15}$  & Argonne &
  intm.  & 3.85 & 4.75  & & 172.2 & 232.8 & & 387.3 & 587.2 & & 396.1 & 594.3 \\
   & & &
  large  & 4.39 & 5.44  & & 196.4 & 264.9 & & 461.1 & 699.6 & & 476.2 & 717.8 \\
   & & CD-Bonn &
  intm.  & 4.15 & 5.11  & & 269.4 & 351.1 & & 339.7 & 514.6 & & 408.1 & 611.7 \\
   & & &
  large  & 4.69 & 5.82  & & 317.3 & 411.5 & & 392.8 & 595.6 & & 482.7 & 727.6 \\
   & & &
     & & & & & & & & & & &    \\
$^{82}Se\rightarrow {^{82}Kr}$  & $3.53~10^{-14}$ & Argonne &
  intm.  & 3.59 &  4.54  & & 164.8 & 225.7 & & 374.5 & 574.2 & & 379.3 & 577.9 \\
   & & &
  large  & 4.18 &  5.29 & & 193.1 & 262.9 & & 454.9 & 697.7 & & 465.1 & 710.2 \\
   & & CD-Bonn &
  intm.  & 3.86 &  4.88 & & 258.7 & 340.4 & & 328.7 & 503.7 & & 390.4 & 594.5 \\
   & & &
  large  & 4.48 &  5.66 & & 312.4 & 408.4 & & 388.0 & 594.4 & & 471.8 & 719.9 \\
   & & &
     & & & & & & & & & & &    \\
$^{100}Mo\rightarrow {^{100}Ru}$  & $5.73~10^{-14}$ &  Argonne &
  intm.  & 3.62 & 4.39 & & 184.9 & 249.8 & & 412.0 & 629.4 & & 405.1 & 612.1 \\
   & & &
  large  & 3.91 & 4.79 & & 191.8 & 259.8 & & 450.4 & 690.3 & & 449.0 & 682.6 \\
   & & CD-Bonn &
  intm.  & 3.96 & 4.81 & & 298.6 & 388.4 & & 356.3 & 543.7 & & 415.9 & 627.9 \\
   & & &
  large  & 4.20 & 5.15 & & 310.5 & 404.3 & & 384.4 & 588.6 & & 454.8 & 690.5 \\
   & & &
     & & & & & & & & & & &    \\
$^{130}Te\rightarrow {^{130}Xe}$ & $5.54~10^{-14}$ &  Argonne &
  intm. &  3.29 & 4.16 & & 171.6 & 234.1 & & 385.1 & 595.2 & & 382.2 & 588.9 \\
   & & &
  large  & 3.34 & 4.18 & & 176.5 & 239.7 & & 405.5 & 626.0 & & 403.1 & 620.4 \\
   & & CD-Bonn &
  intm.  & 3.64 & 4.62 & & 276.8 & 364.3 & & 335.8 & 518.8 & & 396.8 & 611.1 \\
   & & &
  large  & 3.74 & 4.70 & & 293.8 & 384.5 & & 350.1 & 540.3 & & 416.3 & 640.7 \\
   & & &
     & & & & & & & & & & &    \\
$^{136}Xe\rightarrow {^{136}Ba}$ & $5.92~10^{-14}$ &  Argonne &
  intm. &  2.30 & 2.29 & & 119.2 & 163.5 & & 275.0 & 425.3 & & 270.5 & 417.2 \\
   & & &
  large  & 2.19 & 2.75 & & 117.1 & 159.7 & & 276.7 & 428.0 & & 271.0 & 418.0 \\
   & & CD-Bonn &
  intm.  & 2.32 & 2.95 & & 121.4 & 166.7 & & 274.4 & 424.3 & & 267.4 & 412.1 \\
   & & &
  large  & 2.61 & 3.36 & & 125.4 & 172.1 & & 297.2 & 460.0 & & 297.0 & 458.8 \\
   & & &
     & & & & & & & & & & &    \\
\hline\hline
\end{tabular}}
  \end{center}
\end{table*}
\noindent potentials and for $g_A =1.25;~1.00$ (i.e., altogether four NMEs).

\section{$\betabeta$-Decay Induced by Two Mechanisms}

  The observation of \betabeta-decay of several
different isotopes is crucial for obtaining information
about the mechanism or mechanisms that
induce the decay. In the analysis we are going to perform
we will employ the lower bound obtained by the EXO collaboration
on the $\betabeta$-decay half-life of $^{136}Xe$ \cite{Auger:2012ar}:
\be
T_{1/2}^{0\nu}(^{136}Xe) > 1.6 \times 10^{25}
\rm{ y\quad \, (90\,\% \,CL)}.
\label{EXO1}
\ee
We use also the lower limits on the \betabeta-decay half-lives of
$^{76}Ge$, $^{82}Se$ and $^{100}Mo$, and of $^{130}Te$
reported by the Heidelberg-Moscow \cite{Baudis:1999xd},  NEMO3
\cite{Barabash:2010bd} and CUORICINO \cite{CUORI} experiments,
respectively, as well as as well the $^{76}Ge$ half-life
reported in \cite{KlapdorKleingrothaus:2006ff} (see also 
\cite{Klap04}):
\be
\begin{split} T^{0\nu}_{1/2}(^{76}Ge) > 1.9\times 10^{25} \rm{ y}~
\cite{Baudis:1999xd},&\quad
T^{0\nu}_{1/2}(^{82}Se) >  3.6\times 10^{23} \rm{ y}~
\cite{Barabash:2010bd},\\
T^{0\nu}_{1/2}(^{100}Mo) >  1.1\times 10^{24} \rm{ y}~
\cite{Barabash:2010bd},&\quad
T^{0\nu}_{1/2}(^{130}Te)
> 3.0\times 10^{24} \rm{ y}~\cite{CUORI}\,.\\
T^{0\nu}_{1/2}(^{76}Ge)= 2.23^{+0.44}_{-0.31}\times 10^{25}\rm{ y}~
\cite{KlapdorKleingrothaus:2006ff}\,.
\end{split}
\label{limit}
\ee
%

Following \cite{FMPSV0311}, we will consider two cases:
\begin{enumerate}
 \item \betabeta-decay induced by two mechanisms whose
interference term in the $\betabeta$-decay rate
is negligible
\footnote{This possibility is realized when, e.g., the
electron currents (responsible for the emission of the
two electrons in the final state), associated with
the two mechanisms considered, have opposite
chiralities.} \cite{HPR83};
 \item \betabeta-decay triggered by two
CP non-conserving mechanisms whose interference term
cannot be neglected.
\end{enumerate}

In the case 1, given the two mechanisms A and B,
the inverse of the $\betabeta$-decay
half-life for a given isotope $(A_i,Z_i)$ reads:
\be
\frac{1}{T_{i} G_i}= |\eta_A|^2 |{M'}^{0\nu}_{i,A}|^2 +
|\eta_B|^2|{M'}^{0\nu}_{i,B}|^2 \,,
\label{hl}
\ee
%
where the index $i$ denotes the isotope.
The values of the phase
space factor $G^{0\nu}_i(E, Z)$, and of the NMEs ${M'}^{0\nu}_{i, A}$
and ${M'}^{0\nu}_{i,B}$ for the mechanisms we will consider
and for the isotopes  $^{76}$Ge, $^{82}Se$, $^{100}$Mo,
$^{130}$Te and $^{136}$Xe of interest,
are listed in Table \ref{table.1}. The LNV
parameters  are defined in section \ref{LNVmechanisms}.
If the two mechanisms A and B inducing the decay are
interfering and CP non-conserving (case 2), 
the inverse of the $\betabeta$-decay half-life of 
the isotope $(A_i,Z_i)$ can be written as:
\be
\frac{1}{T^{0\nu}_{1/2,i}G^{0\nu}_i(E, Z) }=|\eta_A|^2
|{M'}^{0\nu}_{i,A}|^2 + |\eta_{B}|^2|{M'}^{0\nu}_{i,B}|^2 +
2\cos\alpha
|{M'}^{0\nu}_{i,A}||{M'}^{0\nu}_{i,B}||\eta_A||\eta_{B}|\,.
\label{hlint}
\ee
%
Here  $\alpha$ is the relative phase of $\eta_{A}$ and $\eta_B$.

 If the $\betabeta$-decay is caused by two non-interfering mechanisms,
the LNV parameters $|\eta_{A}|^2$ and $|\eta_B|^2$ characterizing the
mechanisms, can be determined, in principle, from data
on the $\betabeta$-decay half-lives of two isotopes, i.e.,
by solving a system of two linear equations.
In the case of two interfering CP non-conserving mechanisms,
the values of the two parameters
$|\eta_{A}|^2$ and $|\eta_B|^2$ and of the cosine
of the relative phase $\alpha$ of $\eta_{A}$ and $\eta_B$
can be obtained from data on the
 $\betabeta$-decay half-lives of three isotopes, i.e.,
by solving a system of three linear equations.

 As was noticed and discussed in detail in \cite{FMPSV0311}, a very important
role in identifying the physical solutions for
$|\eta_{A}|^2$ and $|\eta_B|^2$ of the corresponding systems of two or
three equations is played by the ``positivity conditions''
$|\eta_{A}|^2 \geq 0$ and $|\eta_B|^2 \geq 0$.
If one of the two mechanisms inducing the $\betabeta$-decay
is the ``standard'' light neutrino exchange,
additional important constraint on the positive solutions
of the relevant systems of two or three
linear equations can be provided by the upper limit on the absolute
neutrino mass scale set by the Moscow and Mainz
$^3H$ $\beta$-decay experiments
\cite{MoscowH3,MainzKATRIN}: $m(\bar{\nu}_e) < 2.3$ eV.
In the case
of $\betabeta$-decay, this limit implies a similar
limit on the effective Majorana mass\footnote{We recall that for $m_{1,2,3} \gtrsim 0.1$ eV
the neutrino mass spectrum is
quasi-degenerate (QD), $m_1\cong m_2\cong m_3 \equiv m$, $m^2_j \gg
\Delta m^2_{21},|\Delta m^2_{31}|$. In this case we have
$m(\bar{\nu}_e) \cong m$ and $\meff \lesssim m$.
}
$\meff < 2.3$ eV.
The latter inequality translates
into the following upper bound on the corresponding
LNV dimensionless parameter $|\eta_\nu|^2 \equiv (\meff/m_e)^2$:
\be
|\eta_\nu|^2\times 10^{12} < 21.2\,. 
\label{etanuH3MM}
\ee
%
The KATRIN    $^3H$ $\beta$-decay experiment
\cite{MainzKATRIN}, which is under preparation,
is planned to have approximately a 10 times better
sensitivity to $m(\bar{\nu}_e)$ than that achieved in
the Moscow and Mainz experiments.
If the designed sensitivity limit of  $\meff < 0.2$ eV (90\% C.L.)
will be obtained in the KATRIN experiment,
it would imply the following rather
stringent upper limit on $|\eta_\nu|^2$:
\be
|\eta_\nu|^2\times 10^{12} < 0.16\,.
\label{etanuKatrin}
\ee
%

 In this work we will derive numerical results using the NMEs
calculated with the large size single particle basis (``large basis'')
and the Argonne potential (``Argonne NMEs''). We report also
results obtained with NMEs calculated with the
Charge Dependent Bonn (CD-Bonn) potential
(``CD-Bonn NMEs'') and compared them
with those derived with the Argonne NMEs.

\subsection{Two Non-interfering Mechanisms}
\label{noninter}

In this case the solutions for the corresponding
two LNV parameters $|\eta_A|^2$ and $|\eta_B|^2$
obtained from data on the $\betabeta$-decay
half-lives of the two isotopes
$(A_i,Z_i)$ and $(A_j,Z_j)$, are given by \cite{FMPSV0311}:
\be
\begin{split}
|\eta_A|^2 =\frac{|{M'}^{0\nu}_{j,B }|^2/T_i G_i- |{M'}^{0\nu}_{i,B
}|^2/T_j G_j} {|{M'}^{0\nu}_{i,A}|^2|{M'}^{0\nu}_{j,B
}|^2-|{M'}^{0\nu}_{i,B}|^2 |{M'}^{0\nu}_{j,A }|^2},\quad
|\eta_B|^2=\frac{ |{M'}^{0\nu}_{i,A }|^2/T_j G_j - |{M'}^{0\nu}_{j,A
}|^2/T_i G_i} {|{M'}^{0\nu}_{i,A}|^2|{M'}^{0\nu}_{j,B }|^2-
|{M'}^{0\nu}_{i,B}|^2|{M'}^{0\nu}_{j,A }|^2}.
\label{solnonint}
\end{split}
\ee
%
It follows from eq. (\ref{solnonint}) that  \cite{FMPSV0311}
if one of the two half-lives, say $T_i$, is fixed,
the positivity conditions $|\eta_A|^2 \geq 0$ and $|\eta_B|^2 \geq 0$
can be satisfied only if $T_j$ lies in a specific ``positivity interval''.
Choosing for convenience always $A_j <A_i$ we get for the positivity
interval \cite{FMPSV0311}:
\be
\frac{ G_i}{G_j}    \frac{  |{M'}^{0\nu}_{i,B }|^2 }{
|{M'}^{0\nu}_{j,B }|^2} T_i \leq T_j \leq \frac{ G_i}{G_j}
\frac{
 |{M'}^{0\nu}_{i,A }|^2 }{ |{M'}^{0\nu}_{j,A }|^2} T_i\,,
\label{PosC}
\ee
%
where we have used
$|{M'}^{0\nu}_{i,A
}|^2/|{M'}^{0\nu}_{j,A }|^2
> |{M'}^{0\nu}_{i,B }|^2 /|{M'}^{0\nu}_{j,B }|^2$.
In the case of
$|{M'}^{0\nu}_{1,A
}|^2/|{M'}^{0\nu}_{2,A }|^2
< |{M'}^{0\nu}_{1,B }|^2 /|{M'}^{0\nu}_{2,B }|^2$,
the interval of values of $T_j$ under discussion
is given by:
\be
\frac{ G_i}{G_j}\,
\frac{
|{M'}^{0\nu}_{i,A }|^2 }{|{M'}^{0\nu}_{j,A }|^2} T_i \leq T_j \leq
\frac{ G_i}{G_j}\, \frac{ |{M'}^{0\nu}_{i,B }|^2 }
{ |{M'}^{0\nu}_{j,B }|^2} T_i\,.
\label{PosC2}
\ee
%

Condition (\ref{PosC}) is fulfilled, for instance, if $A$
is the heavy right-handed (RH) Majorana
neutrino exchange and $B$ is the light Majorana
neutrino exchange  in the case of Argonne NMEs
(see Table \ref{table.1}).
The inequality in eq. (\ref{PosC}) (or (\ref{PosC2}))
 has to be combined
with the experimental lower bounds on the half-lives of
the considered nuclei,  $T^{exp}_{i~min}$.
If, e.g.,  $T^{exp}_{i~min}$ is the lower bound of interest
for the isotope $(A_i,Z_i)$, i.e.,
if  $T_i \geq T^{exp}_{i~min}$, we get from  eq. (\ref{PosC}):
\be
T_j \geq \frac{ G_i}{G_j} \frac{  |{M'}^{0\nu}_{i,B }|^2 }
{ |{M'}^{0\nu}_{j,B }|^2} T^{exp}_{i~min}\,.
\label{lim}
\ee
%
The lower limit in eq. (\ref{lim}) can be larger than the
existing experimental lower bound on $T_j$.
Indeed, suppose that
$T_i\equiv T^{0\nu}_{1/2}(^{136}Xe)$,
$T_j\equiv T^{0\nu}_{1/2}(^{76}Ge)$ and that
the \betabeta-decay is due by the standard light
neutrino exchange and the heavy
RH Majorana neutrino exchange.
In this case the positivity conditions 
for $|\eta_\nu|^2$ and $|\eta_R|^2$ imply
for the  Argonne and CD-Bonn NMEs corresponding to $g_A=1.25\, (1.0)$:
\be
1.90\, (1.85) \leq \frac{T^{0\nu}_{1/2}(^{76}Ge)}{T^{0\nu}_{1/2}(^{136}Xe)}
\leq 2.70\,(2.64), \qquad \rm{(Argonne~NMEs)}\,;
\label{nuRHNAr}
\ee
\be
1.30\, (1.16) \leq \frac{T^{0\nu}_{1/2}(^{76}Ge)}{T^{0\nu}_{1/2}(^{136}Xe)}
\leq 2.47\,(2.30),\qquad \mbox{(CD-Bonn~NMEs)}\,.
\label{nuRHNCDB}
\ee
%
Using the EXO result, eq. (\ref{EXO1}),
and the Argonne NMEs
we get the lower bound on $ T^{0\nu}_{1/2}(^{76}Ge)$:
\be
T^{0\nu}_{1/2}(^{76}Ge)\geq 3.03 \,(2.95)\times 10^{25}\,\rm{y}.
\label{Arg125}
\ee
%
This lower bound is significantly bigger that
the experimental lower bound on $T^{0\nu}_{1/2}(^{76}Ge)$
quoted in eq. (\ref{limit}).
If we use instead the CD-Bonn NMEs,
the limit we obtain is close to the experimental lower bound on
$T^{0\nu}_{1/2}(^{76}Ge)$:
\be
T^{0\nu}_{1/2}(^{76}Ge)\geq  2.08\,(1.85)\times 10^{25}\,\rm{y}.
\label{CDBonn125}
\ee
%

\begin{figure}[h!]
   \begin{center}
 \subfigure
   {\includegraphics[width=7cm]{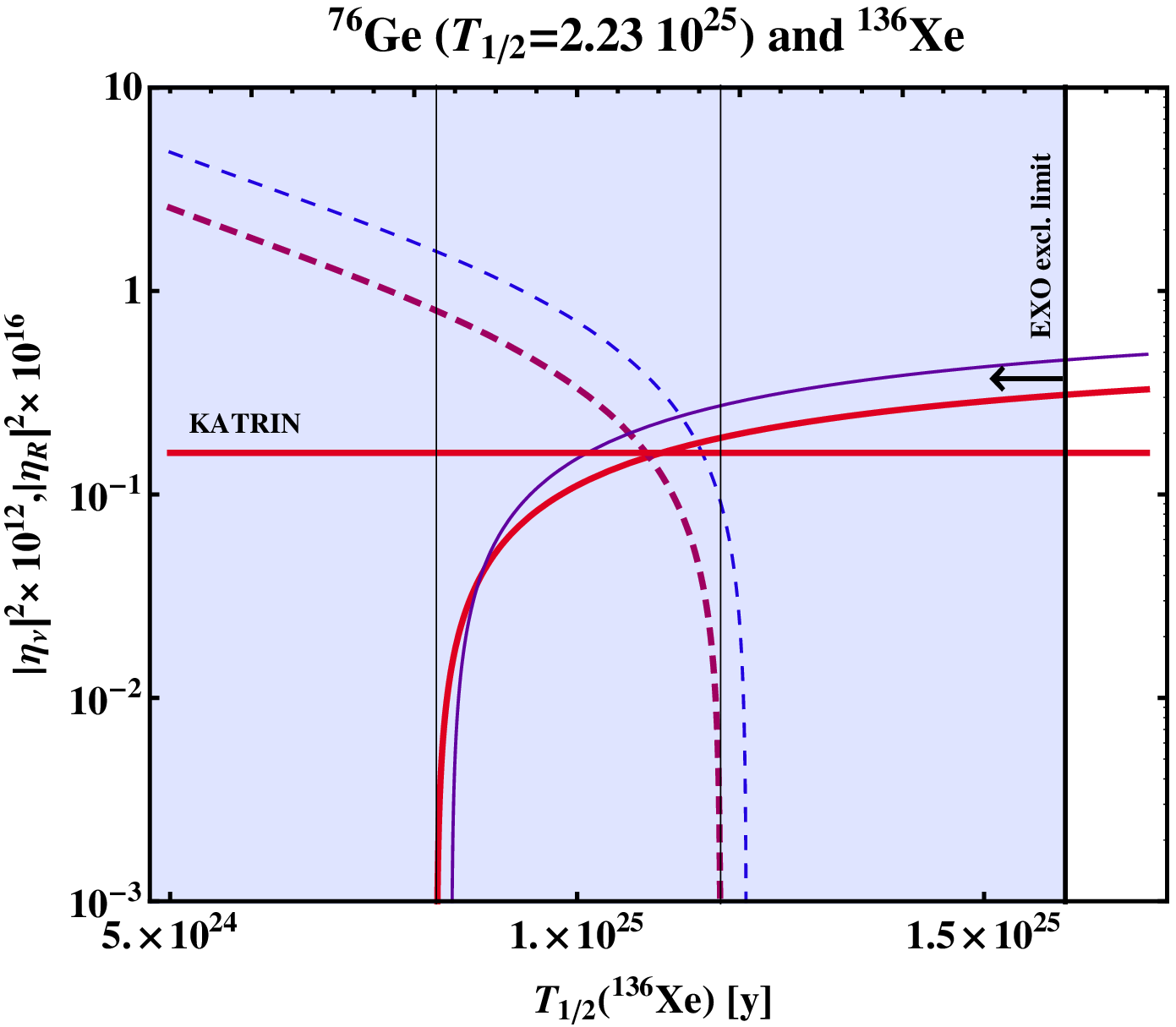}}
 \subfigure
   {\includegraphics[width=7cm]{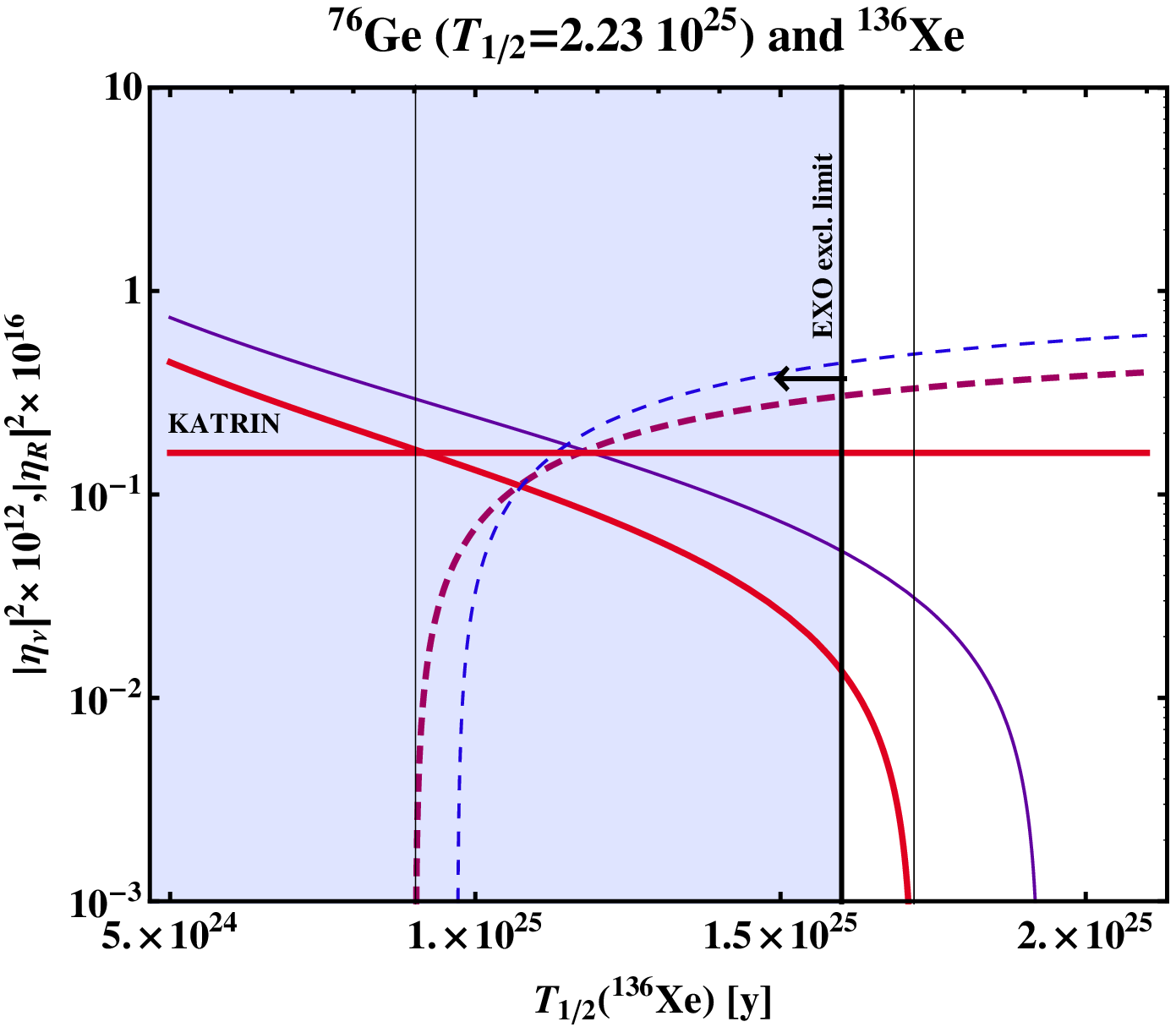}}
     \end{center}
 \vspace{-0.8cm}
 \caption{
\label{fig:figILL}
The values of 
$|\eta_\nu|^2$ (solid lines) and $|\eta_R|^2$ (dashed lines) 
obtained for $T^{0\nu}_{1/2}(^{76}Ge)=2.23 \times 10^{25}$ y
\cite{KlapdorKleingrothaus:2006ff} 
as a function of  $T^{0\nu}_{1/2}(^{136}Xe)$, 
using the Argonne (left panel) and CD-Bonn (right panel) NMEs
corresponding to $g_A=1.25$ (thick lines) and  $g_A=1$
(thin lines). The region of physical (positive) solutions 
for $g_A=1.25$ are delimited by the two vertical lines. 
The solid horizontal line corresponds to the prospective 
upper limit from 
the KATRIN experiment \cite{MainzKATRIN}, while the thick
solid vertical line indicates the EXO lower bound  \cite{Auger:2012ar}.
The gray areas correspond to excluded values of
$|\eta_\nu|^2$ and $|\eta_R|^2$.
}
\end{figure}

For illustrative purposes we  show
in Fig. \ref{fig:figILL} the solutions
of equation (\ref{solnonint}) for $|\eta_\nu|^2$ and $|\eta_R|^2$
derived by fixing  $T^{0\nu}_{1/2}(^{76}Ge)$ to the
best fit value claimed in
\cite{KlapdorKleingrothaus:2006ff},
$T^{0\nu}_{1/2}(^{76}Ge)=2.23 \times 10^{25}$
(see eq. (\ref{limit})).
As Fig. \ref{fig:figILL} shows,
the positive (physical) solutions obtained using
the Argonne NMEs are incompatible with the
EXO result, eq. (\ref{EXO1}), and under the assumptions made
and according to our oversimplified analysis,
are ruled out. At the same time, the physical solutions
obtained using the CD-Bonn NMEs are compatible with
the EXO limit for values of $|\eta_\nu|^2$ and $|\eta_R|^2$
lying in a relatively narrow interval.
%

\begin{figure}[h!]
   \begin{center}
 \subfigure
   {\includegraphics[width=7cm]{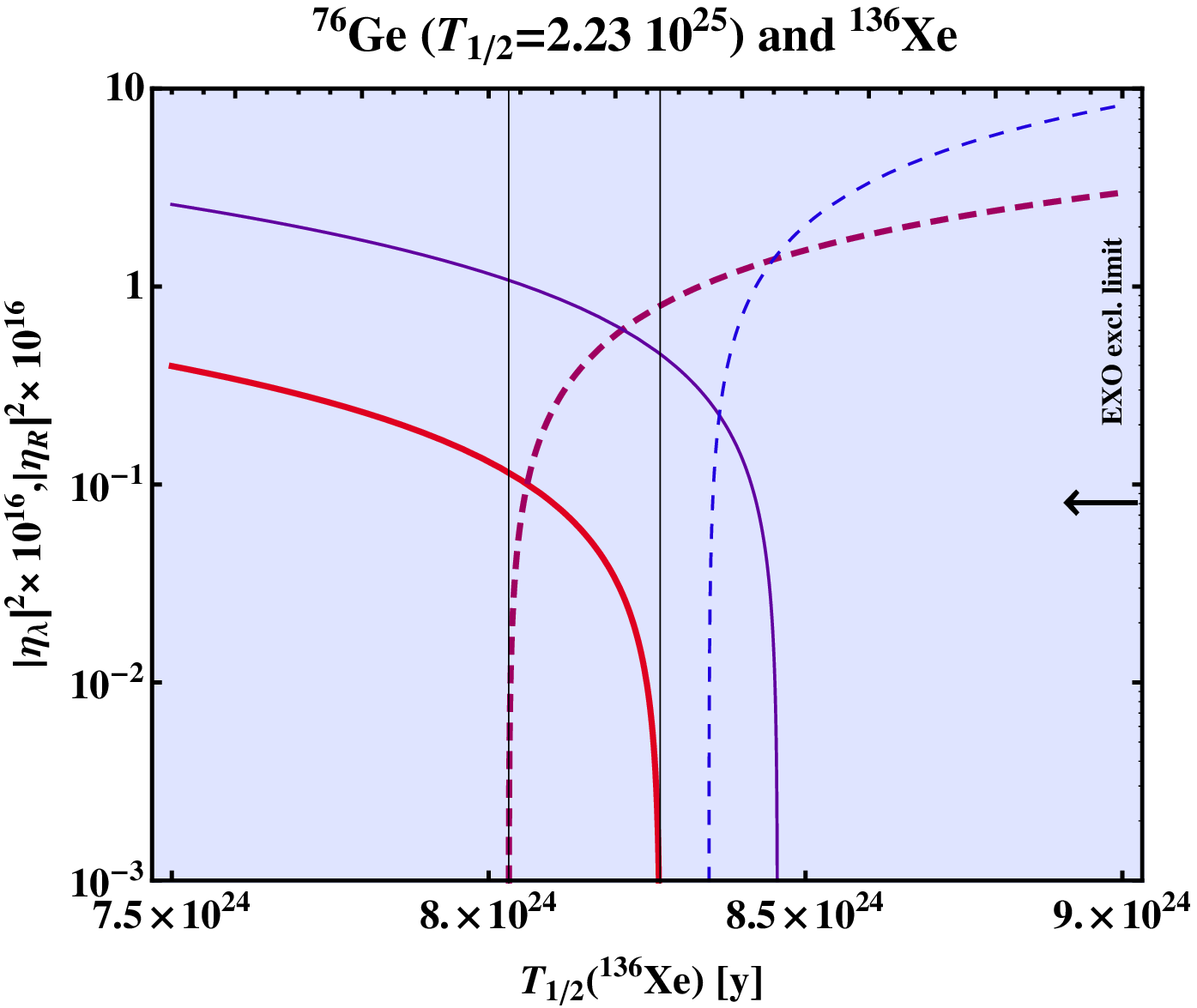}}
 \subfigure
   {\includegraphics[width=7cm]{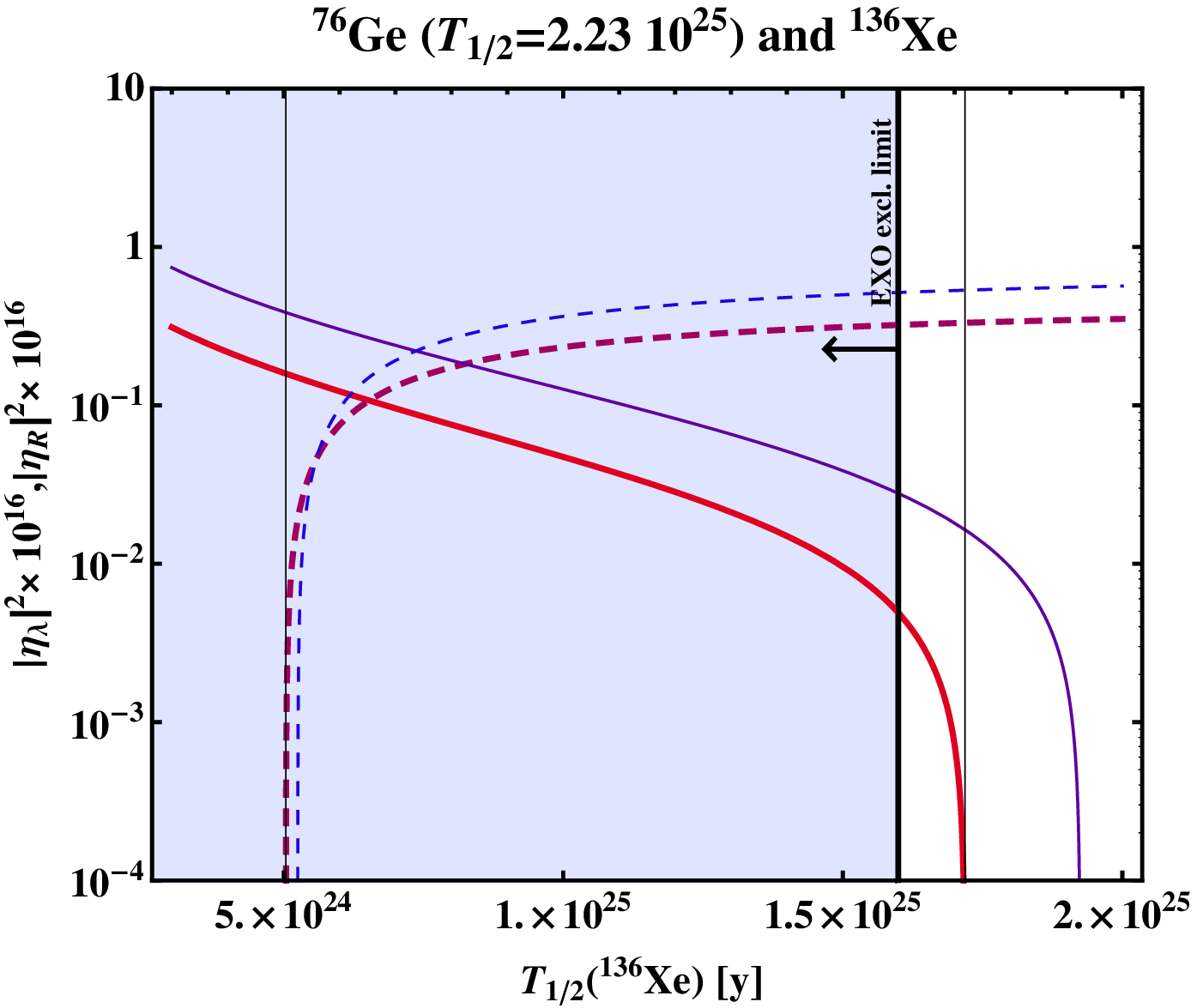}}
     \end{center}
 \vspace{-0.8cm}
 \caption{
\label{fig:figglu}
The same as in Fig. \ref{fig:figILL} but for
the values of the rescaled parameters
$|\eta_{\lambda'}|^2$ (solid lines) and $|\eta_R|^2$ (dashed lines).
}
\end{figure}

We consider next a second example of two non-interfering
$\betabeta$-decay mechanisms, i.e., \betabeta-decay
induced by the heavy
RH  Majorana neutrino exchange and the
gluino exchange. Setting, as above,
$T_i\equiv T^{0\nu}_{1/2}(^{136}Xe)$ and
$T_j\equiv T^{0\nu}_{1/2}(^{76}Ge)$,
we get for the positivity intervals
using the Argonne or CD-Bonn NMEs corresponding to
$g_A=1.25\, (1.0)$:
\be
2.70\, (2.64) \leq \frac{T^{0\nu}_{1/2}(^{76}Ge)}{T^{0\nu}_{1/2}(^{136}Xe)}
\leq 2.78\,(2.67), \qquad \mbox{(Argonne~NMEs)}\,;
\label{RHNgluinoAr}
\ee
\be
1.30 \, (1.16) \leq  \dfrac{T^{0\nu}_{1/2}(^{76}Ge)}{T^{0\nu}_{1/2}(^{136}Xe)}
\leq 4.43\,(4.25), \qquad \mbox{(CD-Bonn~NMEs)}\,,
\label{RHNgluinoCDB}
\ee
%
The lower bound on $ T^{0\nu}_{1/2}(^{76}Ge)$ following
from the EXO limit
in the case of the Argonne NMEs obtained with $g_A=1.25\, (1.0)$ reads:
\be
T^{0\nu}_{1/2}(^{76}Ge)\geq 4.31 \,(4.22)\times 10^{25}\,\rm{y}.
\label{Argon125}
\ee
%
This lower limit is by a factor of 2.27
bigger than the experimental lower limit quoted in
eq. (\ref{limit}). It is also incompatible with the
$4\sigma$ range of values of $T^{0\nu}_{1/2}(^{76}Ge)$
found in \cite{KlapdorKleingrothaus:2006ff}.
The lower bound obtained using the  CD-Bonn NMEs is less stringent:
\be
T^{0\nu}_{1/2}(^{76}Ge)\geq  2.08\,(1.85)\times 10^{25}\,\rm{y}.
\ee
%
In Fig. \ref{fig:figglu} we show that if
$T^{0\nu}_{1/2}(^{76}Ge)=2.23 \times 10^{25}$, the recent EXO lower limit
allows  positive (physical) solutions
for the corresponding two LNV parameters
$|\eta_{\lambda'}|^2$ and $|\eta_R|^2$
only for the  CD-Bonn NMEs
and for values of  $|\eta_{\lambda'}|^2$ and $|\eta_R|^2$
lying in a very narrow interval.
%

 We get similar results
for the third pair of non-interfering mechanisms -
the  squark-neutrino exchange and the heavy
RH neutrino exchange. Indeed, using the Argonne NMEs
corresponding to $g_A=1.25\, (1.0)$ we find for
the positivity interval:
\be
2.52\, (2.40) \leq \frac{T^{0\nu}_{1/2}(^{76}Ge)}{T^{0\nu}_{1/2}(^{136}Xe)}
\leq 2.70\,(2.64)\,,
\label{sqnuRHNAr}
\ee
%
The EXO lower bound in this case implies:
\be
T^{0\nu}_{1/2}(^{76}Ge)\geq 4.03 \,(3.84)\times 10^{25}\,\rm{y}.
\ee
%
From the NMEs computed with the CD-Bonn potential we get
\be
1.30 \, (1.16) \leq  \dfrac{T^{0\nu}_{1/2}(^{76}Ge)}{T^{0\nu}_{1/2}(^{136}Xe)}  \leq 2.95\,(2.81)\,,
\label{sqnuRHNCDB}
\ee
%
and
\be
T^{0\nu}_{1/2}(^{76}Ge)\geq  2.08\,(1.85)\times 10^{25}\,\rm{y}.
\ee
%

 In the case the non-interfering LH  and RH heavy Majorana
neutrino exchanges, the
NMEs for the two mechanisms coincide
and the system of equation in (\ref{solnonint})
reduces to a relation between the half-lives
of the two considered isotopes:
\be
T_j = T_i \frac{G_i|{M'}^{0\nu}_{i,N}|^2 }{G_j|{M'}^{0\nu}_{j,N}|^2 }\,.
\label{LHNRHN}
\ee
%
In this case the EXO lower bound implies the following
lower limits on  $T_j\equiv T^{0\nu}_{1/2}(^{76}Ge)$
for the sets of NMEs we are considering for
$g_A=1.25$ (1.0):
\be
T^{0\nu}_{1/2}(^{76}Ge)\geq 4.31\,(4.22)\times 10^{25}\,\rm{y}
\qquad \mbox{(Argonne~NMEs)}\,,
\ee
\be
T^{0\nu}_{1/2}(^{76}Ge)\geq 2.08\,(1.85)\times 10^{25}\,\rm{y}
\qquad \mbox{(CD-Bonn~NMEs)}.
\ee
%

The range of positive solutions for the LNV
parameters in  equation (\ref{PosC}) shifts towards larger
values if $T_i$ is increased. As
we noticed in \cite{FMPSV0311}, if the experimentally
determined interval of allowed values of the ratio $T_j/T_i$
of the half-lives of the two isotopes considered,
including all relevant uncertainties, lies outside the range of
positive solutions for $|\eta_A|^2$ and
$|\eta_B|^2$, one would be led to conclude that
the $\betabeta$-decay is not generated
by the two mechanisms under discussion.

 Assuming the half-lives of two isotopes, say,
of $^{76}Ge$ and $^{136}Xe$,
$T^{0\nu}_{1/2}(^{76}Ge)\equiv T_1$ and
$T^{0\nu}_{1/2}(^{136}Xe) \equiv T_2$,
to be known, and $\betabeta$-decay triggered by
a pair of non-interfering mechanisms $A$ and $B$,
one can always use the physical solutions
for $|\eta^{LFV}_{A, B}|^2(T_1, T_2)$,
obtained using the two half-lives $T_{1,2}$
(in eq. (\ref{solnonint})), to find the range
of the half-life of a third isotope:
\be
\frac{1}{T_{3} }= G_3(|\eta_A(T_1, T_2)|^2 |{M'}^{0\nu}_{3,A}|^2
+ |\eta_B(T_1, T_2)|^2|{M'}^{0\nu}_{3,B}|^2)\,,
 \label{3pred}
\ee
%
In Tables \ref{tab:prediction2} and \ref{tab:prediction3}
we give numerical predictions based on this observation.
Fixing the half-life of  $^{76}Ge$ to $T_1=10^{26}$ y and
assuming the $^{136}Xe$ half-life $T_2$ lies in an
interval compatible with the existing constraints,
the system of two equations is solved and
the values of  $|\eta_A|^2> 0$ and $|\eta_B|^2> 0$ thus obtained are
used to get predictions for the half-life of a third isotope, in
this case $^{82}Se$, $^{100}Mo$ and $^{130}Te$.
The mechanisms considered are a) light and heavy RH Majorana neutrino
exchanges (Table \ref{tab:prediction2}) and b) gluino and
heavy RH Majorana neutrino exchanges (Table \ref{tab:prediction3}).
It follows from the results shown in Tables  \ref{tab:prediction2}
and \ref{tab:prediction3} that the intervals of allowed values of
the half-lives of  $^{82}Se$, $^{100}Mo$ and $^{130}Te$
thus obtained i) are rather narrow
\footnote{We note that the experimental lower bounds
quoted in eq. (\ref{limit}) have to be
taken into account since, in principle,
they can further constrain the range of
allowed values of $|\eta_A|^2$ and $|\eta_B|^2$ and
of the half-life of the third isotope of interest.},
and ii) exhibit weak dependence on the
NMEs used to derive them (within the sets of
NMEs considered).
\begin{table}
\centering \caption{\label{tab:prediction2}
Predictions using Argonne and CD-Bonn NMEs
corresponding to  $g_A$=1.25 ($g_A$=1 in parenthesis)
in the case of two non-interfering mechanism:
light and heavy RH Majorana neutrino exchanges.
The physical solutions for $|\eta_{\nu}|^2$ and
$|\eta_{R}|^2$ derived for given half-lives of
$^{76}Ge$ and $^{136}Xe$, are used to obtain
predictions for the half-lives of
$^{82}Se$, $^{100}Mo$ and $^{130}Te$.
The $^{76}Ge$ half-life was set to $T(^{76}Ge)=10^{26}$ yr,
while the interval of values of the  $^{136}Xe$ half-life
was determine from the positivity conditions.
}
\vspace{5pt}
\renewcommand{\tabcolsep}{1.1mm}
 \renewcommand{\arraystretch}{1.2}
{\footnotesize
\begin{tabular}{cc}
\hline
\multicolumn{2}{c}{\textbf{Argonne NMEs}} \\
\hline
 Positive solutions &  Predictions\\
\hline
& $ 2.30 (2.34)\cdot 10^{25}<T(^{82}Se)< 2.39(2.49)\cdot 10^{25}$\\
 $ 3.71(3.79)\cdot 10^{25}< T(^{136}Xe) < 5.27(5.42)\cdot 10^{25}$
    & $  1.45(1.46)\cdot10^{25}< T(^{100}Mo) < 1.80(1.76)\cdot 10^{25}$\\
 & $ 1.76(1.78)\cdot 10^{25}< T(^{130}Te) < 2.44(2.49)\cdot 10^{25}$\\\hline
\multicolumn{2}{c}{\textbf{CD-Bonn NMEs}} \\
\hline
 Positive solutions &  Predictions\\
\hline
 & $ 2.30 (2.33)\cdot 10^{25}<T(^{82}Se)< 2.39(2.48)\cdot 10^{25}$\\
 $ 4.04(4.35)\cdot 10^{25}< T(^{136}Xe) < 7.71(8.63)\cdot 10^{25}$
    & $  1.44(1.45)\cdot10^{25}< T(^{100}Mo) < 1.78(1.74)\cdot 10^{25}$\\
  & $ 1.65(1.68)\cdot 10^{25}< T(^{130}Te) < 2.21(2.27)\cdot 10^{25}$\\
\hline
\end{tabular}}
\end{table}
\begin{table}
\centering \caption{\label{tab:prediction3}
The same as in Table \ref{tab:prediction2}
but for the gluino and RH heavy
Majorana neutrino exchange mechanisms.
}\vspace{5pt}
\renewcommand{\tabcolsep}{1.1mm}
 \renewcommand{\arraystretch}{1.2}
{\footnotesize
\begin{tabular}{cc}
\hline \multicolumn{2}{c}{\textbf{Argonne NMEs}} \\
\hline
 Positive solutions &  Predictions\\
\hline

 & $ 2.27 (2.32)\cdot 10^{25}<T(^{82}Se)< 2.30(2.34)\cdot 10^{25}$\\
 $ 3.60(3.74)\cdot 10^{25}< T(^{136}Xe) < 3.71(3.79)\cdot 10^{25}$
    & $  1.43(1.459)\cdot10^{25}< T(^{100}Mo) < 1.45(1.460)\cdot 10^{25}$\\
& $ 1.76(1.78)\cdot 10^{25}< T(^{130}Te) < 1.80(1.86)\cdot 10^{25}$\\\hline
\multicolumn{2}{c}{\textbf{CD-Bonn NMEs}} \\
\hline
Positive solutions &  Predictions\\
\hline
 & $ 2.27 (2.32)\cdot 10^{25}<T(^{82}Se)< 2.30(2.33)\cdot 10^{25}$\\
 $ 2.26(2.35)\cdot 10^{25}< T(^{136}Xe) < 7.71(8.63)\cdot 10^{25}$
    & $  1.43(1.4542)\cdot10^{25}< T(^{100}Mo) < 1.44(1.4543)\cdot 10^{25}$\\
  & $ 1.65(1.68)\cdot 10^{25}< T(^{130}Te) < 1.75(1.81)\cdot 10^{25}$\\
\hline
\end{tabular}}
\end{table}

 One can use eq. (\ref{3pred}) and the lower bound, e.g.,
on $T^{0\nu}_{1/2}(^{136}Xe)$ reported  by
the EXO experiment, to derive a lower bound
on one of the half-lives involved in the study of two
non-interfering mechanisms, say $T_1$.
Indeed, we can set  $T_3= T^{0\nu}_{1/2}(^{136}Xe)$,
in eq. (\ref{3pred}),
use the explicit form of the
solutions for $|\eta^{LFV}_{A, B}|^2(T_1, T_2)$
and apply the existing EXO lower bound.
We get:
\be
\frac 1 T_3 = \frac{D_1}{N T_1} +\frac{D_2}{N T_2}<
\frac{1}{1.6\times 10^{25}\,{\rm y}},
\label{T3limit}
\ee
%
where
\be
D_1= \frac{G_3}{G_1} \left(|{M'}^{0\nu}_{2,A }|^2
|{M'}^{0\nu}_{3,B }|^2- |{M'}^{0\nu}_{3,A }|^2 |{M'}^{0\nu}_{2,B
}|^2 \right), \quad
D_2= \frac{G_3}{G_2}  \left(|{M'}^{0\nu}_{3,A }|^2
|{M'}^{0\nu}_{1,B }|^2- |{M'}^{0\nu}_{1,A }|^2 |{M'}^{0\nu}_{3,B
}|^2 \right),\nn
\ee
%
and
\be
N=|{M'}^{0\nu}_{2,A }|^2 |{M'}^{0\nu}_{1,B }|^2-
|{M'}^{0\nu}_{1,A }|^2 |{M'}^{0\nu}_{2,B }|^2\,.
\ee
%
Using further the positivity constraint given in eq. (\ref{PosC}),
\be
a\, T_1 \leq T_2 \leq b\, T_1\,,
\label{ab}
\ee
%
where $b \equiv |{M'}^{0\nu}_{1,A}|^2/|{M'}^{0\nu}_{2,A}|^2
> a \equiv|{M'}^{0\nu}_{1,B}|^2 /|{M'}^{0\nu}_{2,B}|^2$, we get
\footnote{The inequality in eq. (\ref{ab}) was derived assuming that
$D_2/N>0$. In the case of $D_2/N<0$ one has to
interchange $a$ and $b$ in it.}
the following lower limit on $T_1$ from eq. (\ref{T3limit}):
\be
T_1 \geq  T_3\, \left(
\frac{D_1}{N}+\frac{D_2}{b\,N}\right) >
 1.6\times 10^{25}\,\rm{y}\, \left(
\frac{D_1}{N}+\frac{D_2}{b\,N}\right)\,.
\label{lower}
\ee
%
This  lower bound on $T_1$ depends only on $T_3$ and on the
NMEs $|{M'}^{0\nu}_{i,A }|^2$ and  $|{M'}^{0\nu}_{j,B}|^2$,
$i,j=1,2,3$.
We give examples of predictions based on the eq. (\ref{lower})
in Tables \ref{tab:pred1} and \ref{tab:pred2}.
We notice that, for the NMEs used in the present study,
the EXO lower limit on  $T^{0\nu}_{1/2}(^{136}Xe)$
sets a lower bound on the half-lives of the other
isotopes considered by us that usually exceed their
respective current experimental lower bounds.
\begin{table}[h!]
\centering
\caption{ \label{tab:pred1}
Lower bound on $T_1$ from eq. (\ref{lower})
using the EXO limit on $T^{0\nu}_{1/2}(^{136}Xe)$,
eq. (\ref{EXO1}), and the Argonne and CD-Bonn NMEs
corresponding to $g_A$=1.25 ($g_A$=1), in the case of
two non-interfering mechanisms -
light and heavy RH Majorana neutrino
exchanges. See text for details. }
\vspace{10pt}
\begin{tabular}{c|c|c}
\hline
 ($T_1$,$T_2$) &  Argonne g$_A$=1.25 (1.0) & CD-Bonn g$_A$=1.25 (1.0)\\
\hline
$^{130}$Te - $^{76}$Ge &  T($^{130}$Te)> 7.40 (7.35)$\cdot$ 10$^{24}$ &   T($^{130}$Te)>  3.43 (3.11) $\cdot$ 10$^{24}$\\
$^{100}$Mo - $^{76}$Ge &  T($^{100}$Mo)> 5.45 (5.19)$\cdot$ 10$^{24}$ &   T($^{130}$Te)>  3.00 (2.70) $\cdot$ 10$^{24}$\\
$^{82}$Se - $^{76}$Ge &  T($^{82}$Se)> 7.25 (7.37)$\cdot$ 10$^{24}$ &   T($^{82}$Se)>  4.76 (4.32) $\cdot$ 10$^{24}$\\
\hline
\end{tabular}
\end{table}
\begin{table}[h!]
\centering
\caption{ \label{tab:pred2}
The same as in Table \ref{tab:pred1} for
the gluino and heavy RH Majorana neutrino
exchange mechanisms. See text for details.}
\vspace{10pt}
\begin{tabular}{c|c|c}
\hline
 ($T_1$,$T_2$) &  Argonne g$_A$=1.25 (1.0) & CD-Bonn g$_A$=1.25 (1.0)\\
\hline
$^{130}$Te - $^{76}$Ge &  T($^{130}$Te)> 7.59 (7.53)$\cdot$ 10$^{24}$ &   T($^{130}$Te)>  3.43 (3.11) $\cdot$ 10$^{24}$\\
$^{100}$Mo - $^{76}$Ge &  T($^{100}$Mo)> 6.25 (6.16)$\cdot$ 10$^{24}$ &   T($^{130}$Te)>  3.00 (2.70) $\cdot$ 10$^{24}$\\
$^{82}$Se - $^{76}$Ge &  T($^{82}$Se)> 9.90 (9.87)$\cdot$ 10$^{24}$ &   T($^{82}$Se)>  4.76 (4.32) $\cdot$ 10$^{24}$\\
\hline
\end{tabular}
\end{table}

%
\subsection{Discriminating between Different
Pairs of  Non-interfering Mechanisms}
%

The first thing to notice is that, as it follows from Table 1,
for each of the four different mechanisms of $\betabeta$-decay
considered, the relative difference between NMEs of the decays of
$^{76}Ge$, $^{82}Se$, $^{100}Mo$ and $^{130}Te$
does not exceed approximately 10\%:
$({M'}^{0\nu}_{j,X} - {M'}^{0\nu}_{i,X})/(0.5({M'}^{0\nu}_{j,X} + {M'}^{0\nu}_{i,X})) \ltap 0.1$,
where $i\neq j$ = $^{76}Ge$,$^{82}Se$,$^{100}Mo$,$^{130}Te$,
and $X$ denotes any one of
the four mechanisms discussed.
As was shown in \cite{FMPSV0311},
this leads to degeneracies
between the positivity intervals of values of the
ratio of the half-lives of any two given
of the indicated four isotopes,
corresponding to the different pairs of
mechanisms inducing the $\betabeta$-decay.
The degeneracies in question
make it practically impossible to distinguish between the
different pairs of $\betabeta$-decay mechanisms,
considered in  \cite{FMPSV0311} and in the present article,
using data on the half-lives of two or more
of the four nuclei $^{76}Ge$, $^{82}Se$, $^{100}Mo$ and $^{130}Te$.
At the same time, it is possible, in principle,
to exclude them all using data on the half-lives
of at least two of the indicated four nuclei \cite{FMPSV0311}.

\begin{figure}[htbp]
\centering
\begin{center}
 \subfigure
   { \includegraphics[width=0.4 \textwidth]{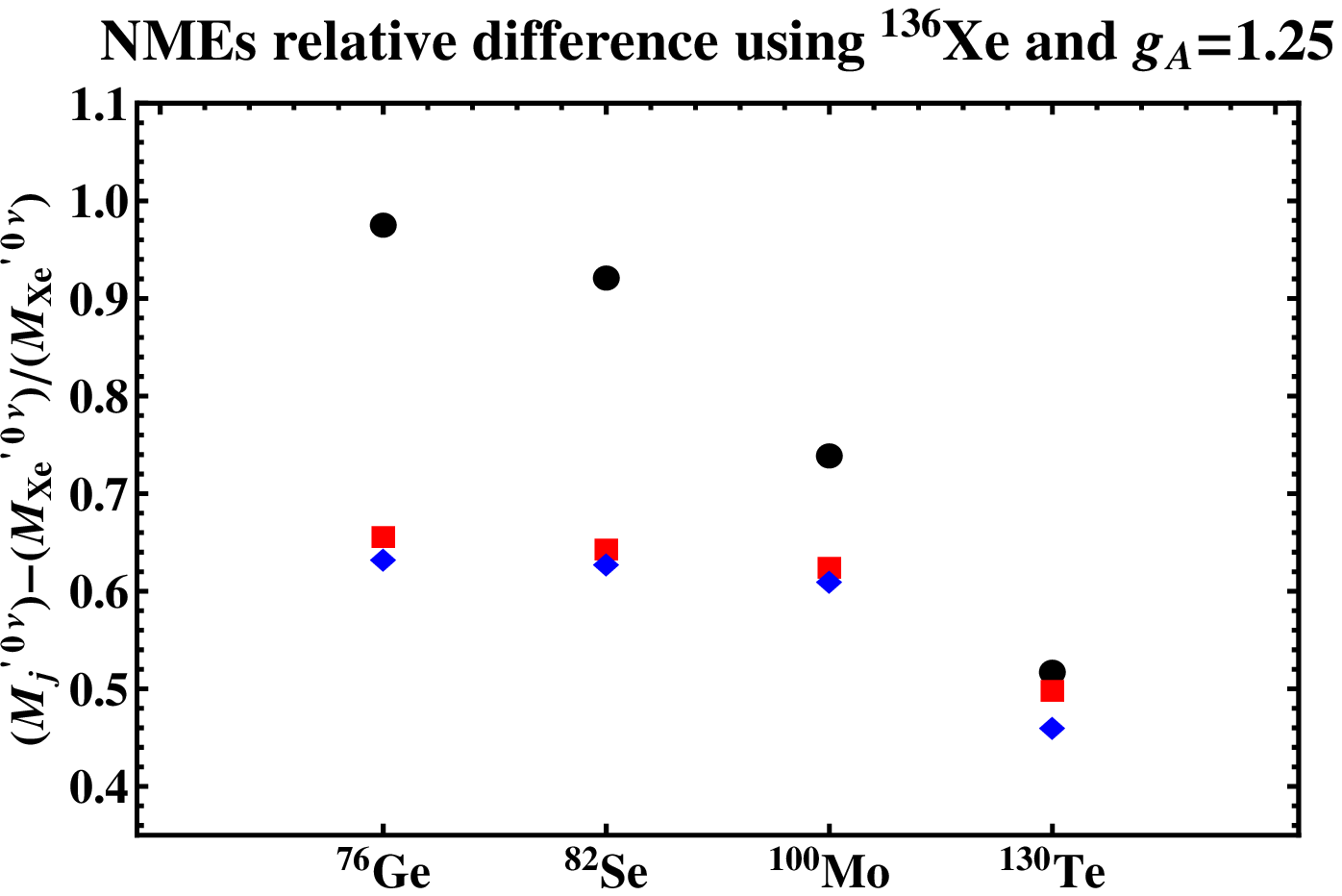}}
 \subfigure
   {\includegraphics[width=0.39 \textwidth]{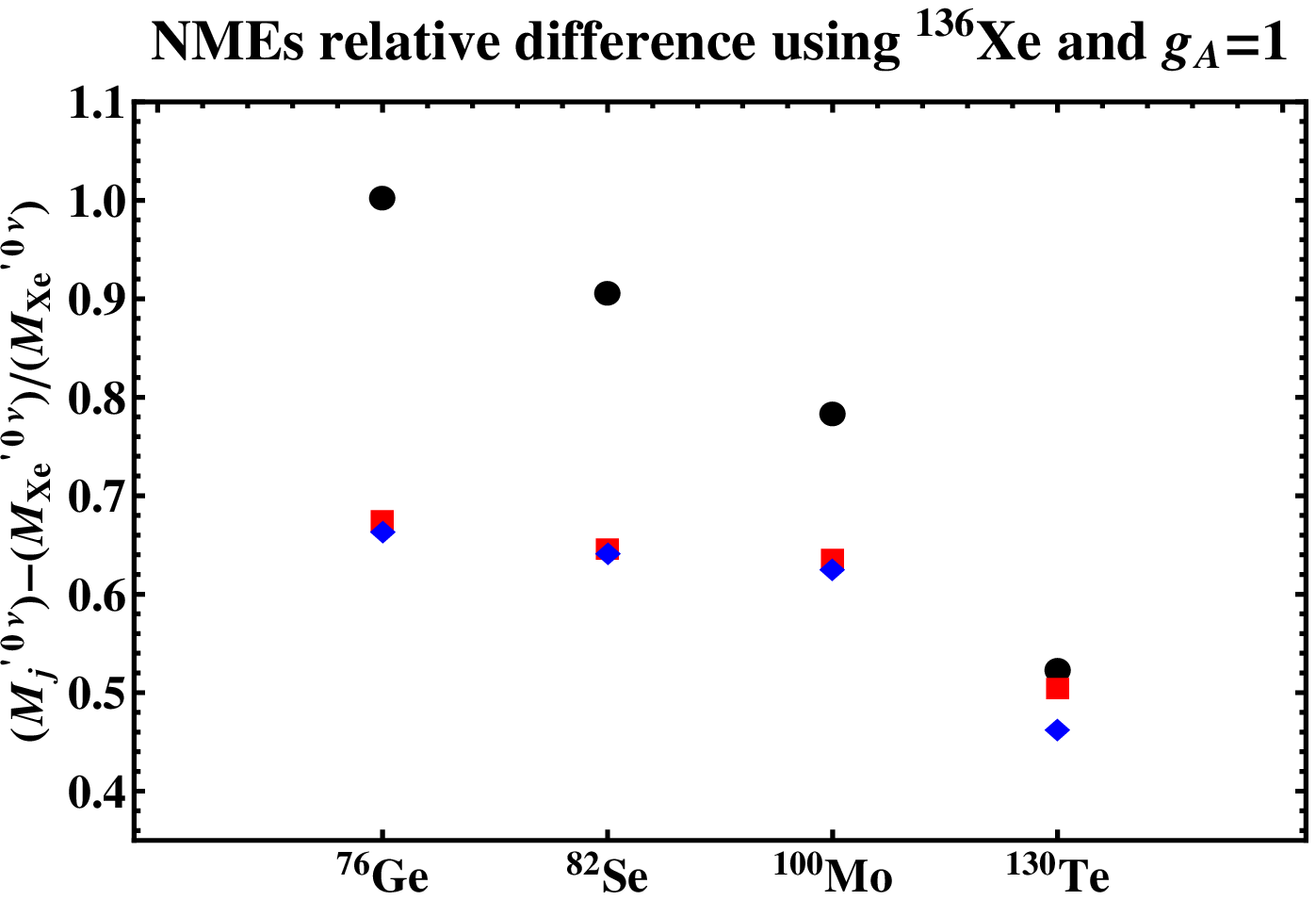}}
   \end{center}
 \vspace{-0.8cm}
   \caption{\label{fig:diff1} The relative differences
between the Argonne  NMEs
$({M'}^{0\nu}_{j}-{M'}^{0\nu}_{i})/{M'}^{0\nu}_{i}$, where $i$=$^{136}Xe$ and
$j$ =$^{76}Ge$,$^{82}Se$,$^{100}Mo$,$^{130}Te$,
for $g_A=1.25$ (left panel) and $g_A=1$ (right panel) and for three different
non-interfering mechanisms: light Majorana neutrino exchange (circles),
RH heavy Majorana neutrino exchange (squares) and gluino exchange
(diamonds). See text for details.}
\end{figure}

\begin{figure}[htbp]
\centering
\begin{center}
 \subfigure
   { \includegraphics[width=0.4 \textwidth]{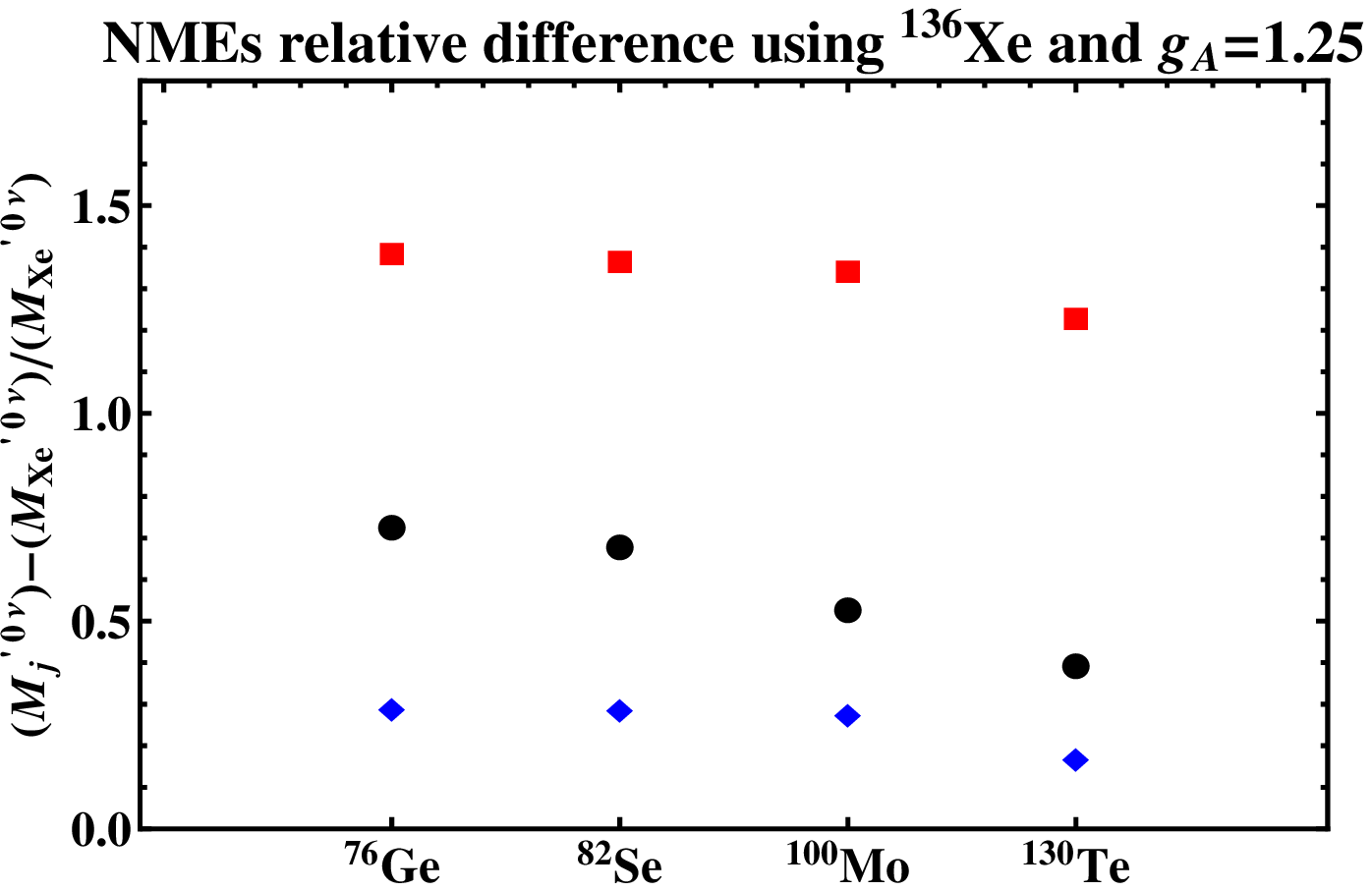}}
 \subfigure
   {\includegraphics[width=0.39 \textwidth]{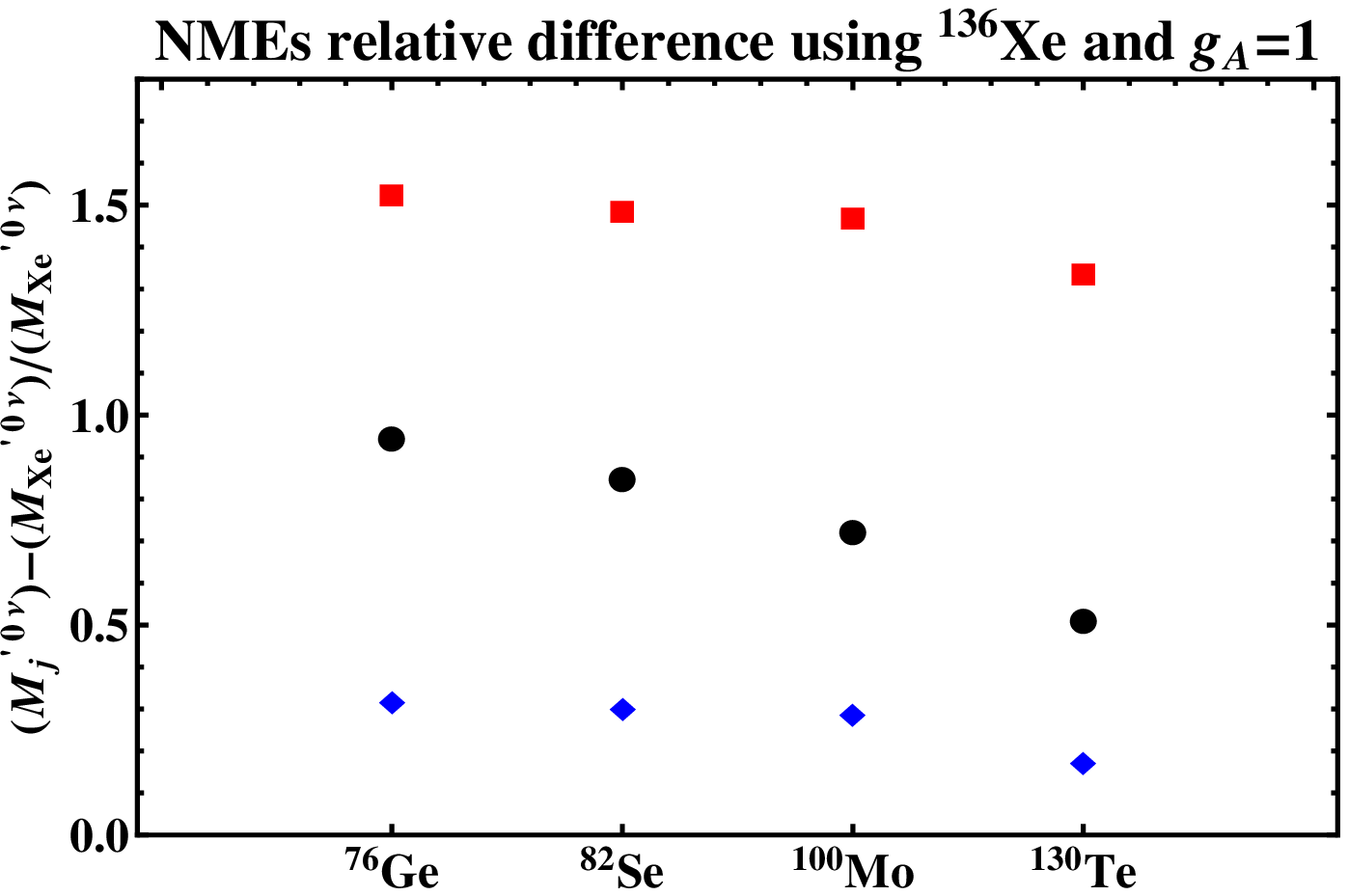}}
   \end{center}
 \vspace{-0.8cm}
   \caption{\label{fig:diff2} The same as in Fig. \ref{fig:diff1}
for the CD-Bonn NMEs. See text for details.}
\end{figure}

 In contrast, the NMEs for the $\betabeta$-decay of
$^{136}Xe$, corresponding to each of the
four different mechanisms we are considering
are by a factor of $\sim (1.3 - 2.5)$
smaller than the $\betabeta$-decay NMEs of the other
four isotopes listed above:
$({M'}^{0\nu}_{j,X} - {M'}^{0\nu}_{i,X})/{M'}^{0\nu}_{i,X}) \cong (0.3-1.5)$,
where $i$ = $^{136}Xe$ and $j$ = $^{76}Ge$,$^{82}Se$,$^{100}Mo$,$^{130}Te$
(see Figs. \ref{fig:diff1}  and \ref{fig:diff2}).
As a consequence,
using data on the half-life of $^{136}Xe$ as input
in determining the positivity interval of values of
the half-life of any second isotope lifts to a certain degree
the degeneracy of the positivity intervals
corresponding to different pairs of non-interfering mechanisms.
This allows, in principle, to draw conclusions about
the pair of mechanisms possibly inducing the $\betabeta$-decay
from data on the half-lives of $^{136}Xe$ and a second
isotope which can be, e.g., any of the four considered
above,  $^{76}Ge$, $^{82}Se$, $^{100}Mo$ and $^{130}Te$.

To be more specific, it follows from eqs. (\ref{nuRHNAr}),
(\ref{RHNgluinoAr}), (\ref{sqnuRHNAr}), (\ref{LHNRHN})
and Table \ref{table.1} that if the  Argonne (CD-Bonn) NMEs derived for
$g_A=1.25\, (1.0)$ are correct,
all four pairs of mechanisms of $\betabeta$-decay
discussed by us will be disfavored, or ruled out,
if it is established experimentally that
$R(^{76}Ge,^{136}Xe) > 2.8~(4.5)$ or that
$R(^{76}Ge,^{136}Xe) < 1.8~(1.1)$, where
$R(^{76}Ge,^{136}Xe)\equiv
T^{0\nu}_{1/2}(^{76}Ge)/T^{0\nu}_{1/2}(^{136}Xe)$.
Further, assuming the validity of the Argonne NMEs,
one would conclude that the light and heavy RH Majorana
neutrino exchanges are the only possible pair of mechanisms
operative in $\betabeta$-decay if it is found experimentally that
$1.9 \leq R(^{76}Ge,^{136}Xe) <2.4$.
For $1.9 \leq R(^{76}Ge,^{136}Xe) <2.6$,
i) the gluino and RH Majorana neutrino exchanges, and
ii)  the LH and RH heavy Majorana neutrino exchanges,
will be disfavored or ruled out.
One finds similar results using the CD-Bonn NMEs.
The numbers we quote in this paragraph should be considered
as illustrative only. In a realistic analysis one has
to take into account the various relevant experimental
and theoretical uncertainties.

  We analyze next the possibility to
discriminate between two pairs of non-interfering mechanisms triggering
the $\betabeta$-decay when
the pairs share one mechanism.
Given three different non-interfering
mechanisms $A$, $B$ and $C$, we can
test the hypothesis of the $\betabeta$-decay induced by the
pairs i) $A+B$ or  ii) $C+B$, using the half-lives of the
same two isotopes.
As  a consequence of the fact that
B is common to both pairs of mechanisms,
the numerators of the expressions for
$|\eta_A|^2$ and $|\eta_C|^2$, as it follows from
eq. (\ref{solnonint}), coincide.
Correspondingly, using the half-lives of the
same two isotopes would allow us to
distinguish, in principle,
between the cases i) and ii)
if the denominators in the expressions for
the solutions for
$|\eta_A|^2$ and  $|\eta_C|^2$
have opposite signs.
Indeed, in this case the physical solutions for $|\eta_A|^2$
in the case i) and $|\eta_C|^2$ in the case ii)  will lie
either in the positivity intervals
(\ref{PosC}) and (\ref{PosC2}), respectively,
or in the intervals (\ref{PosC2}) and (\ref{PosC}).
Thus, the positivity solution intervals for
$|\eta_A|^2$ and $|\eta_C|^2$ would not overlap, except
for the point corresponding to a value of the second isotope
half-life where $\eta_A = \eta_C = 0$.
This would allow, in principle, to discriminate between
the two considered pairs of mechanisms.

 It follows from the preceding discussion that
in order to be possible to
discriminate between the pairs $A+B$ and $C+B$ of
non-interfering mechanisms of $\betabeta$-decay, the following condition
has to be fulfilled:
\be
\frac{Det \begin{pmatrix}
|{M'}^{0\nu}_{i,A}|^2 & |{M'}^{0\nu}_{i,B}|^2 \\
 |{M'}^{0\nu}_{j,A }|^2& |{M'}^{0\nu}_{j,B
}|^2\end{pmatrix}  }{   Det \begin{pmatrix}
|{M'}^{0\nu}_{i,C}|^2 & |{M'}^{0\nu}_{i,B}|^2 \\
 |{M'}^{0\nu}_{j,C }|^2& |{M'}^{0\nu}_{j,B
}|^2\end{pmatrix}    }=\frac{|{M'}^{0\nu}_{i,A}|^2|{M'}^{0\nu}_{j,B
}|^2-|{M'}^{0\nu}_{i,B}|^2 |{M'}^{0\nu}_{j,A
}|^2}{|{M'}^{0\nu}_{i,C}|^2|{M'}^{0\nu}_{j,B
}|^2-|{M'}^{0\nu}_{i,B}|^2 |{M'}^{0\nu}_{j,C }|^2}<0\,.
\ee
%
This condition is satisfied if one of the following two sets of
inequalities holds:
\ba
\label{ineqI}
I)&& \frac{{M'}^{0\nu}_{j,C}-{M'}^{0\nu}_{i,C}}{{M'}^{0\nu}_{i,C}}
< \frac{{M'}^{0\nu}_{j,B}-{M'}^{0\nu}_{i,B}}{{M'}^{0\nu}_{i,B}} <
\frac{{M'}^{0\nu}_{j,A}-{M'}^{0\nu}_{i,A}}{{M'}^{0\nu}_{i,A  }},\\
\label{ineqII}
II)&&  \frac{{M'}^{0\nu}_{j,A}-{M'}^{0\nu}_{i,A}}{{M'}^{0\nu}_{i,A
}}    <
\frac{{M'}^{0\nu}_{j,B}-{M'}^{0\nu}_{i,B}}{{M'}^{0\nu}_{i,B}} <
\frac{{M'}^{0\nu}_{j,C}-{M'}^{0\nu}_{i,C}}{{M'}^{0\nu}_{i,C}}.
 \ea
%

One example of a possible application of the preceding
results is provided by the mechanisms of light Majorana neutrino
exchange (A), RH heavy Majorana neutrino exchange (B) and
gluino exchange (C) and the Argonne NMEs.
We are interested in studying cases involving
$^{136}Xe$ since, as it was already discussed  earlier,
the NMEs of $^{136}Xe$ differ significantly from
those of the lighter isotopes such as $^{76}Ge$
(see Table \ref{table.1}).
Indeed, as can be shown,
it is possible, in principle, to discriminate
between the two pairs $A+B$ and $C+B$ of the three
mechanisms indicated above
if we combine data on the half-life of
$^{136}Xe$ with those on the half-life of one of the four
isotopes $^{76}Ge$, $^{82}Se$, $^{100}Mo$ and $^{130}Te$,
and use the Argonne NMEs in the analysis.
In this case the inequalities (\ref{ineqI}) are realized,
as can be seen in Fig. \ref{fig:diff1},
where we plot the relative differences
$({M'}^{0\nu}_{j}-{M'}^{0\nu}_{i})/{M'}^{0\nu}_{i}$
for the Argonne NMEs where the indices $i$ and $j$ refer
respectively to $^{136}Xe$ and to one of the four isotopes
$^{76}Ge$, $^{82}Se$, $^{100}Mo$ and $^{130}Te$.
In the case of the CD-Bonn NMEs (Fig. \ref{fig:diff2}),
the inequalities (\ref{ineqI}) or (\ref{ineqII})
do not hold for the pairs of mechanisms considered.
The inequalities given in eq. (\ref{ineqI}) hold,
as it follows from Fig. \ref{fig:diff2},
if, e.g., the mechanisms A, B and C are respectively
the heavy RH Majorana neutrino exchange, the light Majorana
neutrino exchange and the gluino exchange.
%

\begin{figure}[htbp]
\centering%
\subfigure
   {\includegraphics[width=7cm]{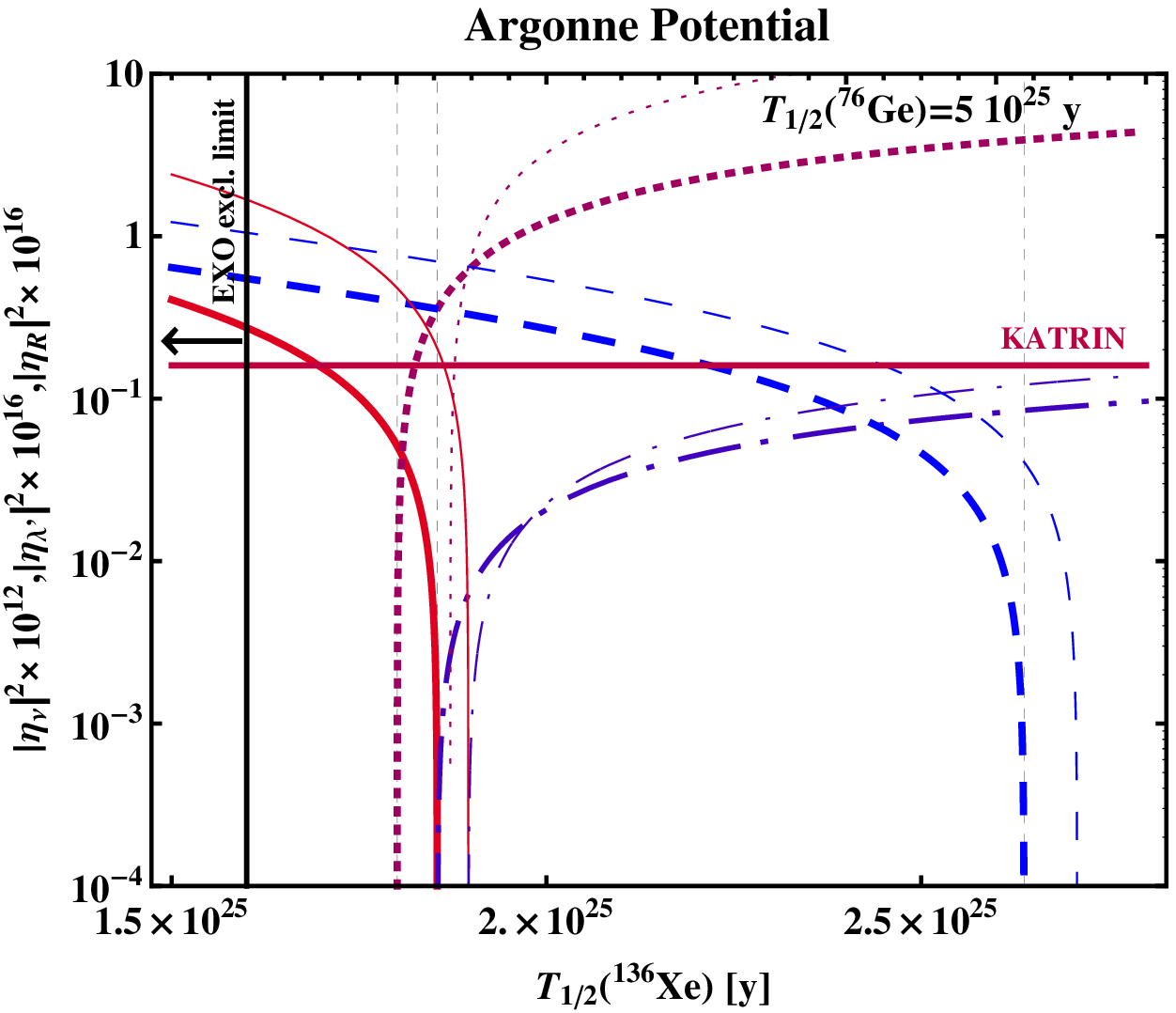}}
 \vspace{2mm}
 \subfigure
   {\includegraphics[width=7cm]{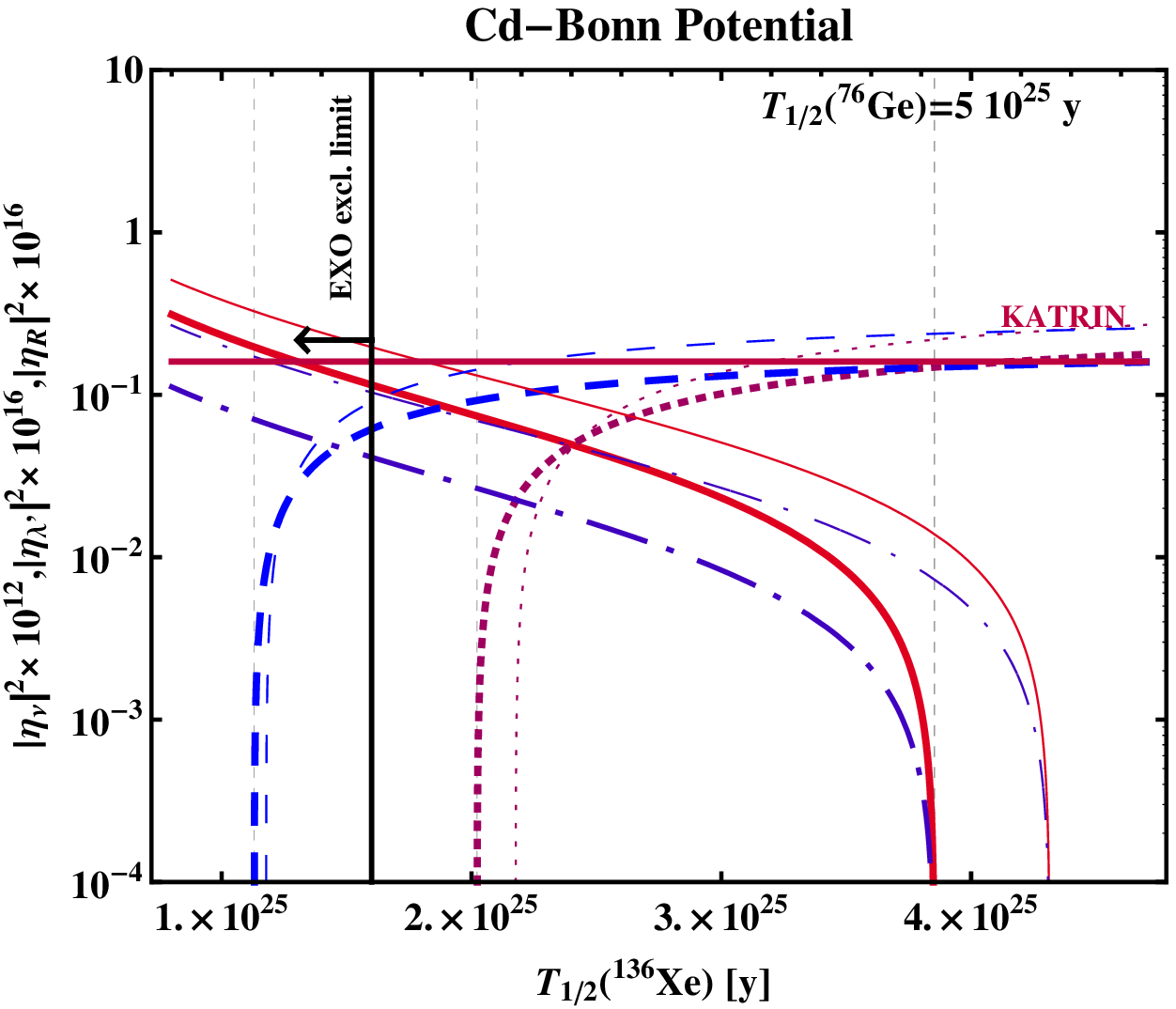}}
 \caption{\label{fig:comparison}
Solutions for the LNV parameters
corresponding to two
pairs of non-interfering mechanisms: i)
$|\eta_\nu|^2$ and $|\eta_R|^2$ (dot-dashed  and dashed lines)  and
ii) $|\eta_{\lambda'}|^2$ and $|\eta_R|^2$ (solid and dotted lines).
The solutions are obtained  by fixing $T_i =
T^{0\nu}_{1/2}(^{76}Ge)=5 \times 10^{25}$ y and letting free $T_j =
T^{0\nu}_{1/2}(^{136}Xe)$ and using the sets of Argonne (left panel)
and CD-Bonn (right panel) NMEs  calculated 
for $g_A= 1.25$ (thick lines) and $g_A=1$ (thin lines).
The range of positive solutions in the case of Argonne NMEs
and $g_A=1.25$ is  delimited by the two vertical dashed lines. The
horizontal solid line corresponds to the prospective upper limit
\meff< 0.2 eV \cite{MainzKATRIN}. The thick solid vertical line
indicates the EXO lower limit on $T^{0\nu}_{1/2}(^{136}Xe)$
\cite{Auger:2012ar}.  See text for details. }
\end{figure}

\begin{figure}[htbp]
\centering%
\subfigure
   {\includegraphics[width=7cm]{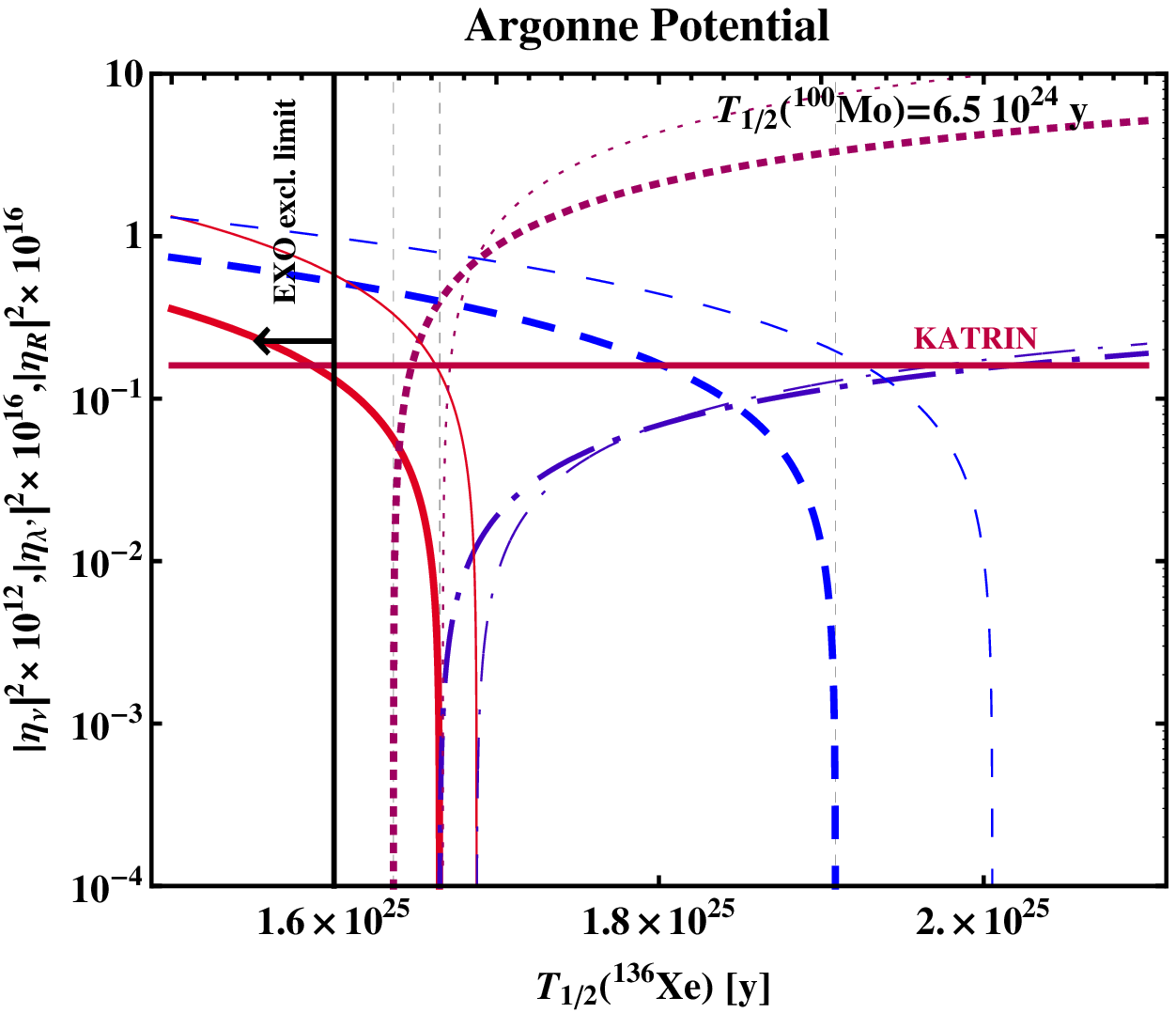}}
 \vspace{2mm}
 \subfigure
   {\includegraphics[width=7cm]{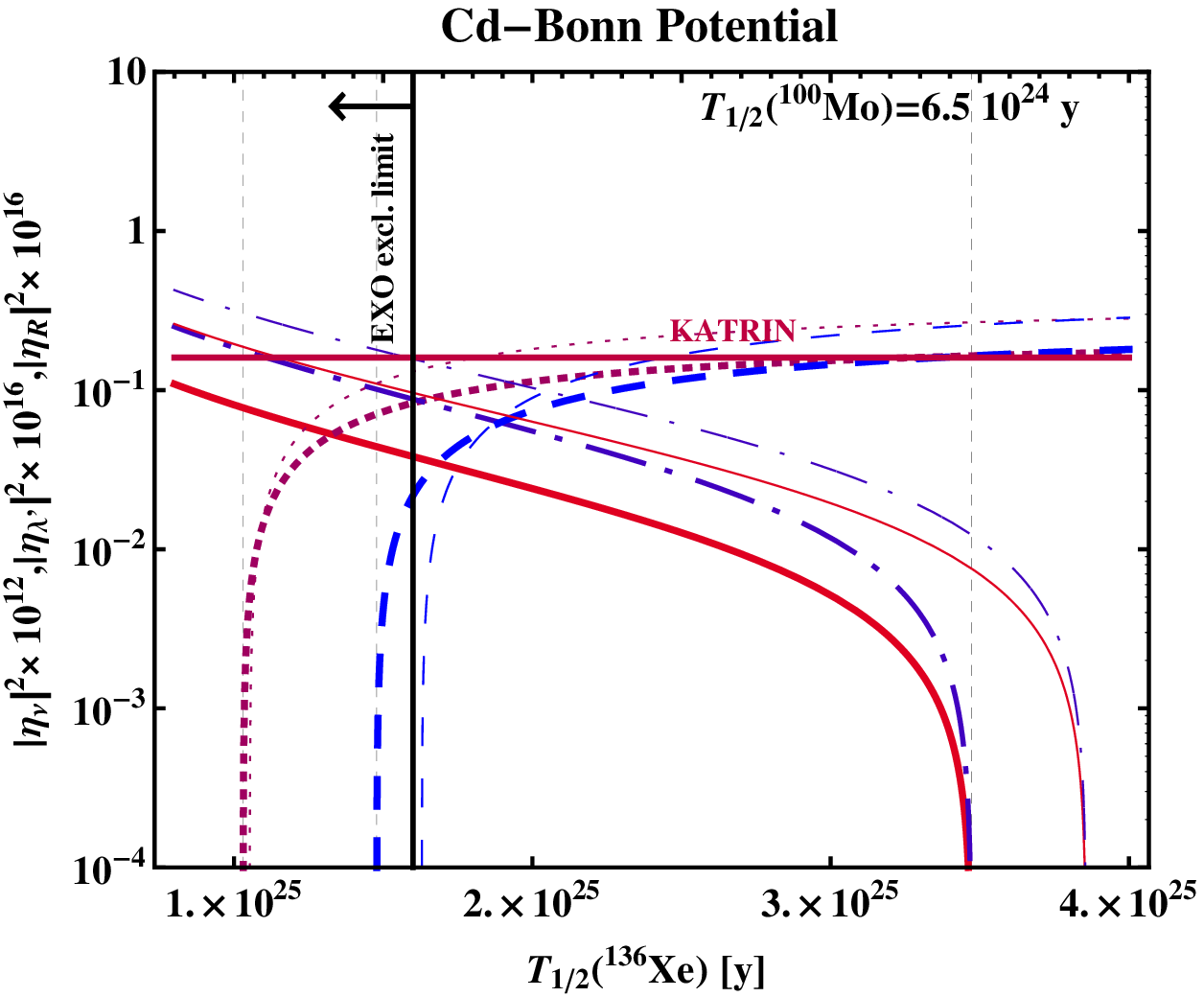}}
 \caption{\label{fig:comparison2}
Solutions for the LNV parameters of  two
pairs of non-interfering $\betabeta$-decay mechanisms i)
$|\eta_\nu|^2$ and $|\eta_R|^2$ (dot-dashed  and dashed lines)  and
ii) $|\eta_{\lambda'}|^2$ and $|\eta_R|^2$
(solid and dotted lines) obtained by fixing $T_i =
T^{0\nu}_{1/2}(^{100}Mo)=6.5 \times 10^{24}$ yr and letting free $T_j =
T^{0\nu}_{1/2}(^{136}Xe)$. The other notations are the same as in
Fig. \ref{fig:comparison}. See text for details. }
\end{figure}

 The preceding considerations are illustrated graphically
in Figs. \ref{fig:comparison} and \ref{fig:comparison2}.
In  Fig. \ref{fig:comparison} we use
$T_i \equiv T^{0\nu}_{1/2}(^{76}Ge)$  and  $T_j \equiv
T^{0\nu}_{1/2}(^{136}Xe)$ and the Argonne (left panel)
and CD-Bonn (right panel) NMEs
for the decays of $^{76}Ge$ and $^{136}Xe$
to show the possibility of discriminating between
the two pairs of non-interfering mechanisms
considered earlier: i) light Majorana neutrino exchange and heavy
RH Majorana neutrino exchange (RHN) and
ii) heavy RH Majorana neutrino exchange and gluino exchange.
The $^{76}Ge$ half-life is set to
$T_i = 5 \times 10^{25}$ y, while that of  $^{136}Xe$, $T_j$,
is allowed to vary in a certain interval.
The solutions for the three LNV parameters
corresponding to the three mechanisms
considered, $|\eta_\nu|^2$, $|\eta_R|^2$ and
$|\eta_{\lambda'}|^2$, obtained for the chosen value
of $T_i$ and interval of values of $T_j$, are shown
as functions of $T_j$. As is clearly seen in the
left panel of Fig. \ref{fig:comparison},
if $|\eta_\nu|^2$, $|\eta_R|^2$ and $|\eta_{\lambda'}|^2$
are obtained using the Argonne NMEs,
the intervals of values of $T_j$ for which one obtains
the physical positive solutions for
$|\eta_\nu|^2$ and $|\eta_{\lambda'}|^2$,
do not overlap. This  makes it possible, in principle,
to determine which of the two pairs of mechanisms considered
(if any) is inducing  the $\betabeta$-decay.
The same result does not hold 
if one uses the CD-Bonn NMEs in the analysis,
as is illustrated in the right panel of
Fig. \ref{fig:comparison}. In this case none of the inequalities
(\ref{ineqI}) and (\ref{ineqII}) is fulfilled,
the intervals of values of $T_j$ for which
one obtains physical solutions for $|\eta_\nu|^2$ and
$|\eta_{\lambda}|^2$ overlap and the discrimination between
the two pairs of mechanisms is problematic.

 We show in Fig. \ref{fig:comparison2} that the
features of the solutions for $|\eta_\nu|^2$ and
$|\eta_{\lambda}|^2$ we have discussed above,
which are related to the
values of the relevant NMEs,
do not change if one uses in the analysis
the half-lives and NMEs of $^{136}Xe$ and
of another lighter isotope instead
of $^{76}Ge$, namely, of $^{100}Mo$.
%
%
\subsection{Two Interfering Mechanisms}
%

  We analyze in the present Section the possibility of
$\betabeta$-decay induced by two interfering
CP-non-conserving mechanisms.
This case is characterized by three parameters:
the absolute values and the relative phase of
the two LNV parameters associated
with the two mechanisms. They can be determined, in principle,
from data on the half-lives of three isotopes, $T_i$, $i=1,2,3$.
Given $T_{1,2,3}$ and denoting by $A$ and
$B$ the two mechanisms, one can set a system of
three linear equations in three unknowns,
the solution of which reads  \cite{FMPSV0311}:
\be
|\eta_A|^2= \frac{D_i}{D}\,,~~ |\eta_{B}|^2 = \frac{D_j}{D}\,,~~
z \equiv 2\cos\alpha|\eta_A||\eta_{B}| = \frac{D_k}{D}\,,
\label{intsol1}
\ee
%
where $D$, $D_i$, $D_j$ and $D_k$ are the following determinants:
\be
D = \left| \begin{array}{ccc}
 ({M'}^{0\nu}_{i,A })^2 & ({M'}^{0\nu }_{i,B })^2 & {M'}^{0\nu }_{i,B } {M'}^{0\nu}_{i,A } \\
 ({M'}^{0\nu}_{j,A })^2 & ({M'}^{0\nu }_{j,B })^2 & {M'}^{0\nu }_{j,B } {M'}^{0\nu}_{j,A }\\
 ({M'}^{0\nu}_{k,A })^2 & ({M'}^{0\nu }_{k,B})^2 & {M'}^{0\nu }_{k,B } {M'}^{0\nu}_{k,A }
\end{array}
\right|\,,~~~
D_i = \left| \begin{array}{ccc}
i/ T_i G_i & ({M'}^{0\nu }_{i,B})^2 & {M'}^{0\nu }_{i,B} {M'}^{0\nu}_{i,A } \\
i/ T_j G_j & ({M'}^{0\nu }_{j,B})^2 & {M'}^{0\nu }_{j,B} {M'}^{0\nu}_{j,A } \\
i/ T_k G_k & ({M'}^{0\nu }_{k,B})^2 & {M'}^{0\nu}_{k,B}
{M'}^{0\nu}_{k,A }
\end{array} \right|\,,
\label{DDi}
\ee
\be
D_j=  \left| \begin{array}{ccc}
({M'}^{0\nu}_{i,A})^2 & i/ T_i G_i &  {M'}^{0\nu }_{i,B} {M'}^{0\nu}_{i,A } \\
({M'}^{0\nu}_{j,A })^2 & i/ T_j G_j &  {M'}^{0\nu }_{j,B} {M'}^{0\nu}_{j,A } \\
({M'}^{0\nu}_{k,A })^2 & i/ T_k G_k &  {M'}^{0\nu }_{k,B}
{M'}^{0\nu}_{k,A }
\end{array}
\right|\,,~~~
D_k = \left| \begin{array}{ccc}
  ({M'}^{0\nu}_{i,A })^2& ({M'}^{0\nu }_{i,B})^2 & i/ T_i G_i \\
  ({M'}^{0\nu}_{j,A })^2& ({M'}^{0\nu }_{j,B})^2 & i/ T_j G_j\\
  ({M'}^{0\nu}_{k,A })^2& ({M'}^{0\nu }_{k,B})^2 & i/ T_k G_k
\end{array} \right|\,.
\label{D2Dk}
\ee
%
As in the case of two non-interfering mechanisms,
the LNV parameters must be non-negative
$|\eta_A|^2 \geq 0 0$ and $|\eta_{B}|^2 \geq 0$, and in addition
the interference term must satisfy the following condition:
\be
-2|\eta_A||\eta_{B}| \leq 2\cos\alpha|\eta_A||\eta_{B}|
\leq2|\eta_A||\eta_{B}|\,.
\label{fase}
\ee
%
These conditions will be called from here
on ``positivity conditions''.

  Using the positivity conditions it is possible
to  determine the interval of positive solutions
for one of the three half-life, e.g., $T_k$,
if the values of the other two half-lives in the equations
have been measured and are known. The condition on the
interference term in equation (\ref{PosC}) can
considerably reduce the interval of values of
$T_k$ where $|\eta_{A}|^2\geq 0$ and $|\eta_{B}|^2\geq 0$.
In Table \ref{tab:tabint} we give examples of the
constraints on $T_k$ following from the positivity conditions
for three different pairs of interfering mechanisms: light Majorana
neutrino and supersymmetric gluino exchange; light Majorana neutrino
exchange and heavy LH Majorana neutrino exchange; gluino
exchange and heavy LH Majorana neutrino exchange.
It follows from the results shown in  Table \ref{tab:tabint}, in
particular, that when T$(^{76}Ge)$ is set to
T$(^{76}Ge)=2.23\times 10^{25};10^{26}$ y,
but T$(^{130}Te)$ is close to the current experimental lower limit,
the positivity constraint intervals of values of  T$(^{136}Xe)$
for the each of the three pairs of interfering mechanisms
considered are incompatible with the
EXO lower bound on  T$(^{136}Xe)$, eq. (\ref{EXO1}).
\begin{table}[h!]
\centering
\caption{ \label{tab:tabint}
Ranges of the half-live of $^{136}Xe$ for different fixed
values of the half-lives of $^{76}Ge$
and $^{130}Te$ in the case of three  pairs
of interfering mechanisms: light Majorana neutrino exchange and gluino
exchange (upper table);
light Majorana and heavy LH Majorana neutrino exchanges
(middle table); gluino exchange and and heavy LH Majorana neutrino
exchange (lower table). The results shown are obtained with
the ``large basis''  $g_A=1.25$  Argonne NMEs.
One star (two stars) indicate that the EXO bound
constrains further (rules out) the corresponding solution.
}
\vspace{15pt}
\begin{tabular}{|l|l|c|}
\hline \hline
 T$^{0\nu}_{1/2}$[y](fixed) &  T$^{0\nu}_{1/2}$[y](fixed) & Allowed Range \\
\hline
  T$(Ge)=2.23\cdot10^{25}$**  &  T$(Te)=3\cdot10^{24}$     &   $ 2.95 \cdot 10^{24}\leq T(Xe) \leq 5.65\cdot 10^{24}$ \\
          T$(Ge)= 10^{26}$**  &  T$(Te)=3\cdot10^{24}$     & $ 3.43\cdot10^{24} \leq T(Xe) \leq   4.66\cdot10^{24}$\\
 T$(Ge)= 2.23\cdot10^{25}$  &  T$(Te)=3\cdot10^{25}$     &$ 1.74\cdot10^{25} \leq T(Xe) \leq1.66 \cdot10^{26}$\\
   T$(Ge)= 10^{26}$         &  T$(Te)=3\cdot10^{25}$     & $2.58 \cdot10^{25} \leq T (Xe) \leq 6.90 \cdot10^{25}$\\
\hline\hline
\end{tabular}\vspace{15pt}
\begin{tabular}{|l|l|c|}
\hline \hline
 T$^{0\nu}_{1/2}$[y](fixed) &  T$^{0\nu}_{1/2}$[y](fixed) & Allowed Range \\
\hline
  T$(Ge)=2.23\cdot10^{25}$**  &  T$(Te)=3\cdot10^{24}$     &   $ 4.93\cdot 10^{24}\leq T(Xe) \leq6.21\cdot 10^{24}$ \\
          T$(Ge)= 10^{26}$** &  T$(Te)=3\cdot10^{24}$     & $5.23 \cdot10^{24} \leq T(Xe) \leq   5.83 \cdot10^{24}$\\
 T$(Ge)= 2.23\cdot10^{25}$  &  T$(Te)=3\cdot10^{25}$     &$ 3.95 \cdot10^{25} \leq T(Xe) \leq 8.25\cdot10^{25}$\\
   T$(Ge)= 10^{26}$         &  T$(Te)=3\cdot10^{25}$     & $4.68 \cdot10^{25} \leq T (Xe) \leq 6.61\cdot10^{25}$\\
\hline\hline
\end{tabular}\vspace{15pt}
\begin{tabular}{|l|l|c|}
\hline \hline
 T$^{0\nu}_{1/2}$[y](fixed) &  T$^{0\nu}_{1/2}$[y](fixed) & Allowed Range \\
\hline
  T$(Ge)=2.23\cdot10^{25}$**  &  T$(Te)=3\cdot10^{24}$     &   $ 5.59\cdot 10^{23}\leq T(Xe) \leq 1.26\cdot 10^{25}$ \\
          T$(Ge)= 10^{26}$*  &  T$(Te)=3\cdot10^{24}$     & $1.21 \cdot10^{24} \leq T(Xe) \leq  4.71 \cdot10^{25}$\\
 T$(Ge)= 2.23\cdot10^{25}$**  &  T$(Te)=3\cdot10^{25}$     &$ 1.05 \cdot10^{24} \leq T(Xe) \leq 2.42 \cdot10^{24}$\\
   T$(Ge)= 10^{26}$*         &  T$(Te)=3\cdot10^{25}$     & $3.32 \cdot10^{24} \leq T (Xe) \leq 2.16\cdot10^{25}$\\
\hline\hline
\end{tabular}
\end{table}
%

 We consider next
a case in which the half-life of $^{136}Xe$ is
one of the two half-lives assumed to have been
experimentally determined.
The \betabeta-decay is supposed to be
triggered by light Majorana neutrino and
gluino exchange mechanisms with
LFV parameters $|\eta_\nu|^2$ and $|\eta_{\lambda'}|^2$.
We use in the analysis the half-lives
of $^{76}Ge$, $^{136}Xe$ and  $^{130}Te$,
which will be denoted for simplicity
respectively as $T_1$, $T_2$ and $T_3$.
Once the experimental bounds on $T_i$,  $i=1,2,3$,
given in eq. (\ref{limit}), are taken into account,
the conditions for destructive interference, i.e.,
for $\cos\alpha <0$, are given by:
\be
 z < 0\,:
\begin{cases}
1.9\times 10^{25} < T_1 \leq 1.90 T_2,  &   T_3 \geq\dfrac{9.64 T_1 T_2}{16.32  T_1 +8.59 T_2}; \\
 1.90 T_2 < T_1 \leq 2.78 T_2, & T_3 >\dfrac{3.82 T_1 T_2}{6.33 T_1 + 3.66 T_2}; \\
 T_1 > 2.78 T_2, & T_3 \geq\dfrac{7.33 T_1 T_2}{11.94 T_1 +7.61 T_2}\,, \\
\end{cases}
\label{destrint}
\ee
%
where we have used the ``large basis''  $g_A=1.25$ Argonne NMEs
(see Table \ref{table.1}).
The conditions for constructive interference read:
\be
 z > 0\,:
\begin{cases}
1.90  T_2< T_1 \leq 2.29 T_2,  &  \dfrac{9.64 T_1 T_2}{16.32  T_1 +8.59 T_2} \leq T_3 \leq  \dfrac{3.82 T_1 T_2}{6.33 T_1 + 3.66 T_2} ;\\
 2.29 T_2< T_1 < 2.78 T_2, & \dfrac{7.33 T_1 T_2}{11.94 T_1 +7.61 T_2} \leq  T_3 \leq \dfrac{3.82 T_1 T_2}{6.33 T_1 + 3.66 T_2}.\\
\end{cases}
\label{constrint}
\ee
%
If we set, e.g.,  the $^{76}Ge$ half-life
to the value claimed in \cite{KlapdorKleingrothaus:2006ff}
$T_1=2.23\times 10^{25}$ y, we find that only destructive
interference between the contributions of the two
mechanisms considered in the $\betabeta$-decay rate, is possible.
Numerically we get in this case
\be
T_3 > \dfrac{3.44 T_2}{5.82  + 1.37 \times 10^{-25} T_2}\,.
\label{destrint2}
\ee
%
For  $1.37 \times 10^{-25} T_2 \ll 5.82$
one finds:
\be
T(^{130}Te) \gtrsim 0.59\, T(^{136}Xe)
\gtrsim 9.46\times 10^{24}~\rm{y}\,,
\label{destrint2v}
\ee
%
where the last inequality has been obtained
using the  EXO lower bound on  $T(^{136}Xe)$.
Constructive interference is possible
for the pair of interfering mechanisms
under discussion only if
$T(^{76}Ge) \gtrsim 3.033 \times 10^{25}$ y.

\begin{figure}[h!]
  \begin{center}
 \subfigure
 {\includegraphics[width=8cm]{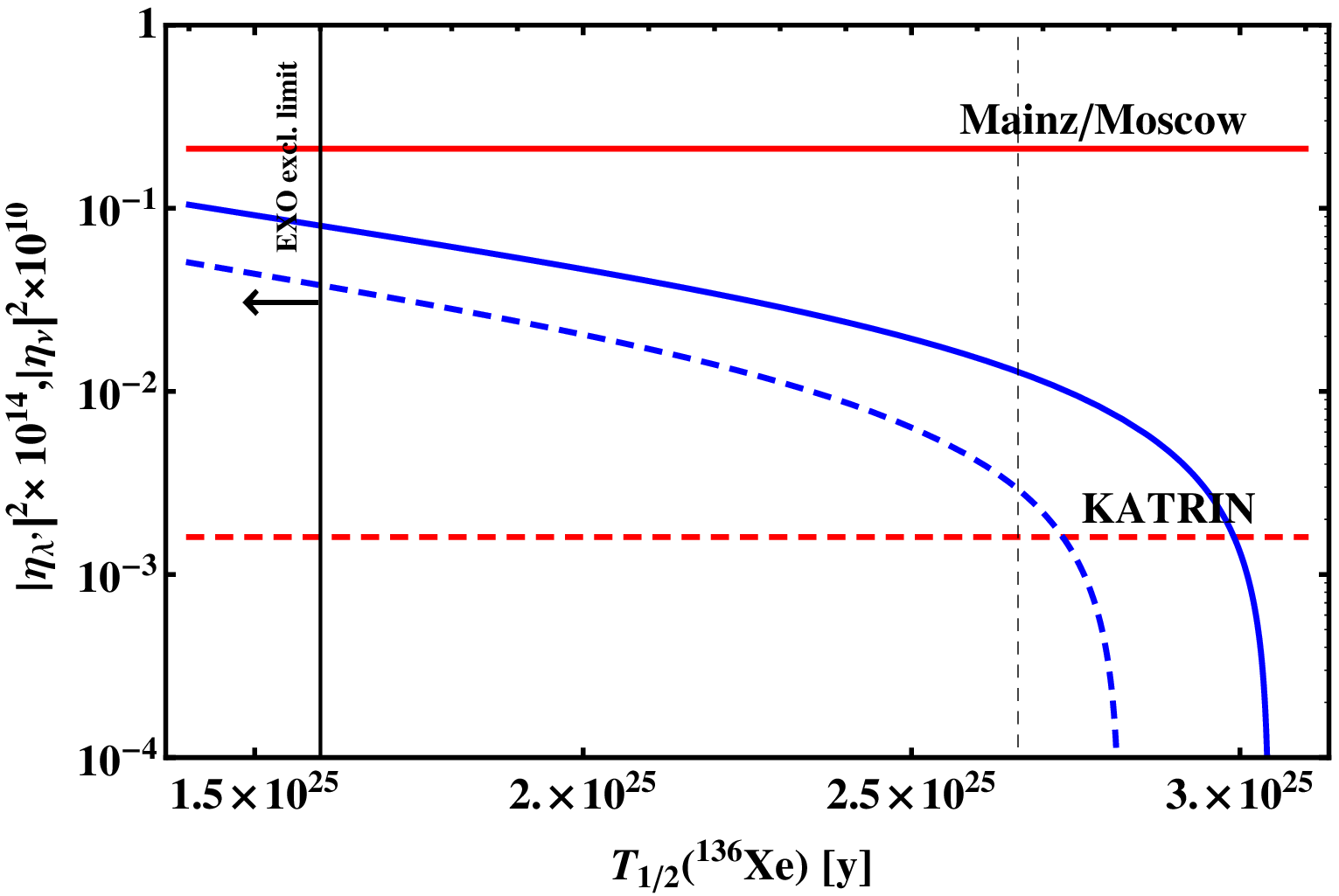}}
 \vspace{5mm}
 \subfigure
   {\includegraphics[width=7.5cm]{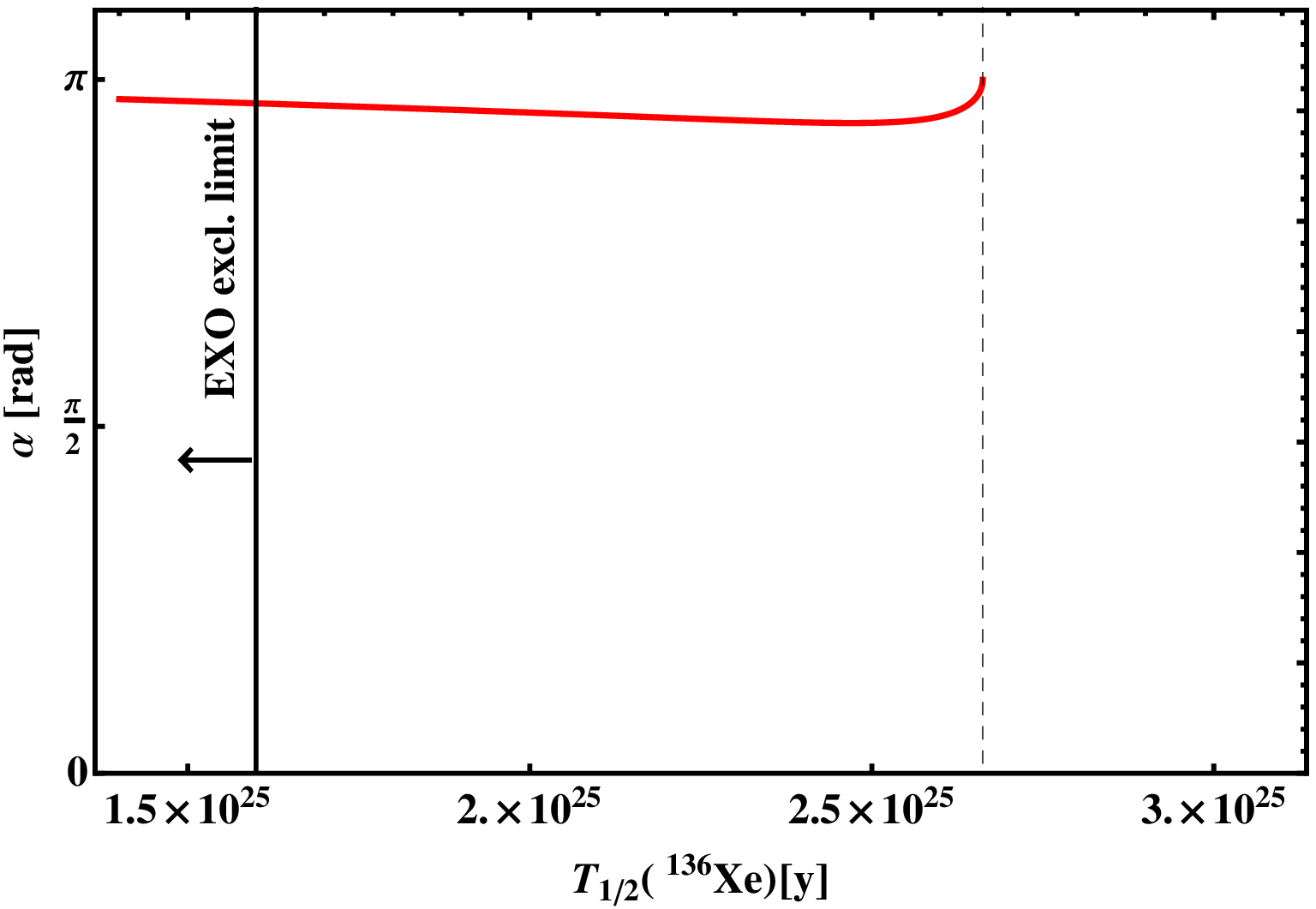}}
     \end{center}
\vspace{-1.0cm}
    \caption{
\label{fig:figInt1}
Left panel: the values of $|\eta_\nu|^2\times 10^{10}$ (thick solid line)
and $|\eta_{\lambda'}|^2\times 10^{14}$ (dotted line),
obtained as solutions of the system of
equations (\ref{hlint}) for fixed values of
$T(^{76}Ge) = 2.23 \times 10^{25}$ y and $T(^{130}Te) = 10^{25}$ y,  and
letting $T^{0\nu}_{1/2}(^{136}Xe)$ free.
The physical allowed regions correspond
to the areas within the two vertical lines.
Right panel:  
the values of the phase $\alpha$ in the allowed interval of
values of $T^{0\nu}_{1/2}(^{136}Xe)$, corresponding to
physical solutions for  $|\eta_\nu|^2$
and $|\eta_{\lambda'}|^2$. In this case  $\cos \alpha < 0$
and the interference is destructive.
See text for details.}
\end{figure}

\begin{figure}[h!]
  \begin{center}
 \subfigure
 {\includegraphics[width=8cm]{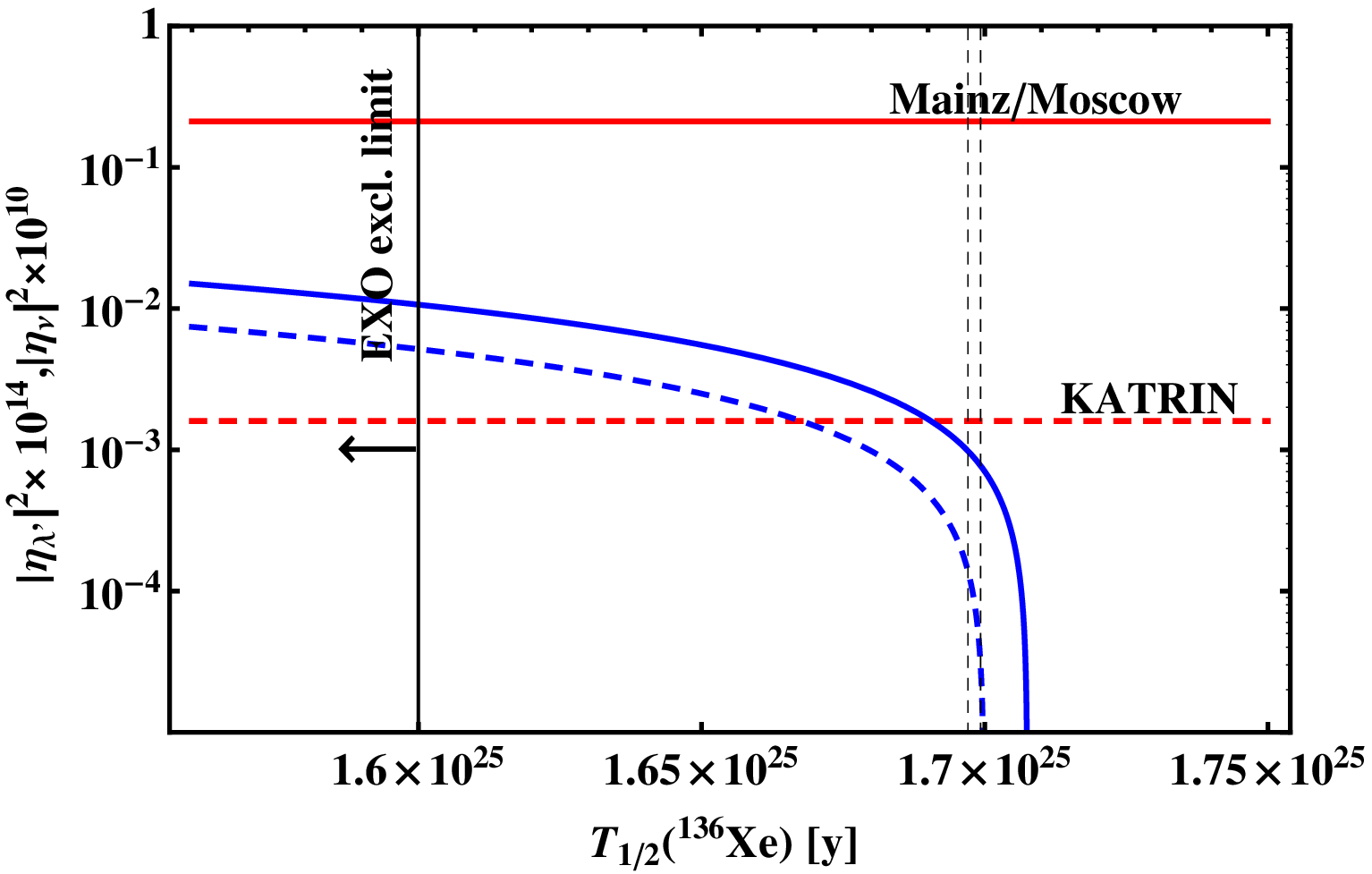}}
 \vspace{5mm}
 \subfigure
   {\includegraphics[width=7.5cm]{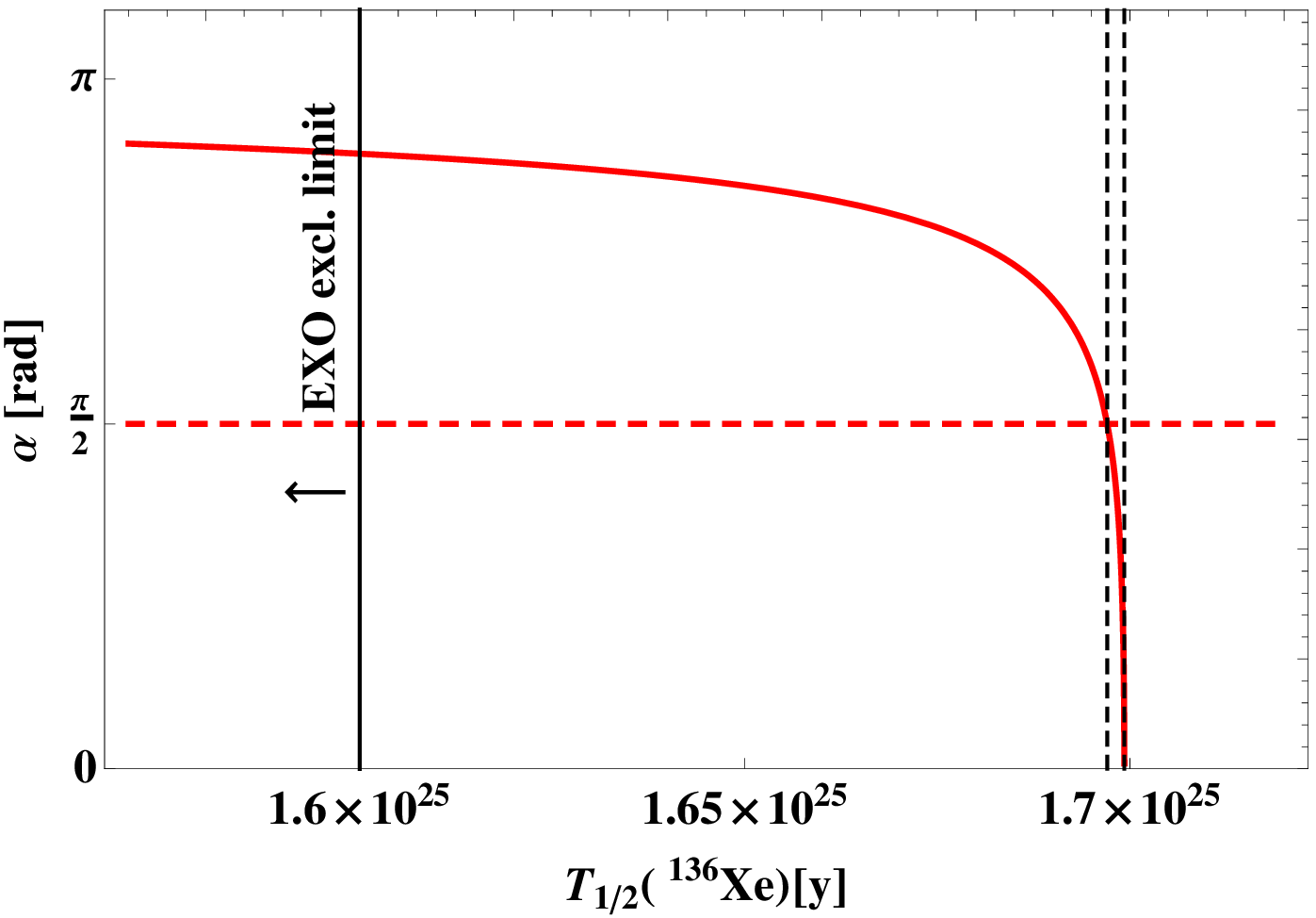}}
     \end{center}
\vspace{-1.0cm}
    \caption{
\label{fig:figInt2}
Left panel:
the same as in Fig. \ref{fig:figInt1} but for
$T(^{76}Ge) = 3.5 \times 10^{25}$ y and $T(^{130}Te) = 8.0 \times 10^{25}$ y.
The interval of values of
$T^{0\nu}_{1/2}(^{130}Xe)$ between
i) the vertical solid and right dashed lines
ii) the two vertical dashed lines,
and iii) the vertical solid and left dashed lines,
correspond respectively to
i) physical (non-negative) solutions for $|\eta_\nu|^2$ and
$|\eta_{\lambda'}|^2$,  ii) constructive interference ($z > 0$),
and iii) destructive interference ($z < 0$).
Right panel: the corresponding values of the phase
$\alpha$ as a function of  $T^{0\nu}_{1/2}(^{130}Xe)$.
Constructive interference is possible
only for values of $T^{0\nu}_{1/2}(^{130}Xe)$
between the two vertical dashed
lines. See text for details.}
\end{figure}

  The possibilities of destructive and constructive interference
are illustrated in Figs.  \ref{fig:figInt1} and \ref{fig:figInt2},
respectively.
In these figures the physical allowed regions, determined through
the positivity conditions, correspond to the areas within the two
vertical lines (the solutions must be compatible also with the
existing lower limits given in eq (\ref{limit})).
For instance, using the Argonne ``large basis''
NMEs corresponding to $g_A = 1.25$
and setting $T(^{76}Ge) = 2.23 \times 10^{25}$ y and $T(^{130}Te) = 10^{25}$ y,
positive solutions are allowed only in the  interval
$1.60 \times 10^{25}\leq  T(^{136}Xe) \leq 2.66 \times 10^{25}$ y
(Fig. \ref{fig:figInt1}).
As can be seen in  Figs.  \ref{fig:figInt1} and \ref{fig:figInt2},
a constructive interference is possible only if $T_2 \equiv
T(^{136}Xe)$ lies in a relatively narrow interval and $T_3\equiv
T(^{130}Te)$ is determined through the conditions in  eq. (\ref{constrint}).
%

\begin{figure}
  \begin{center}
 \subfigure
 {\includegraphics[width=7.5cm]{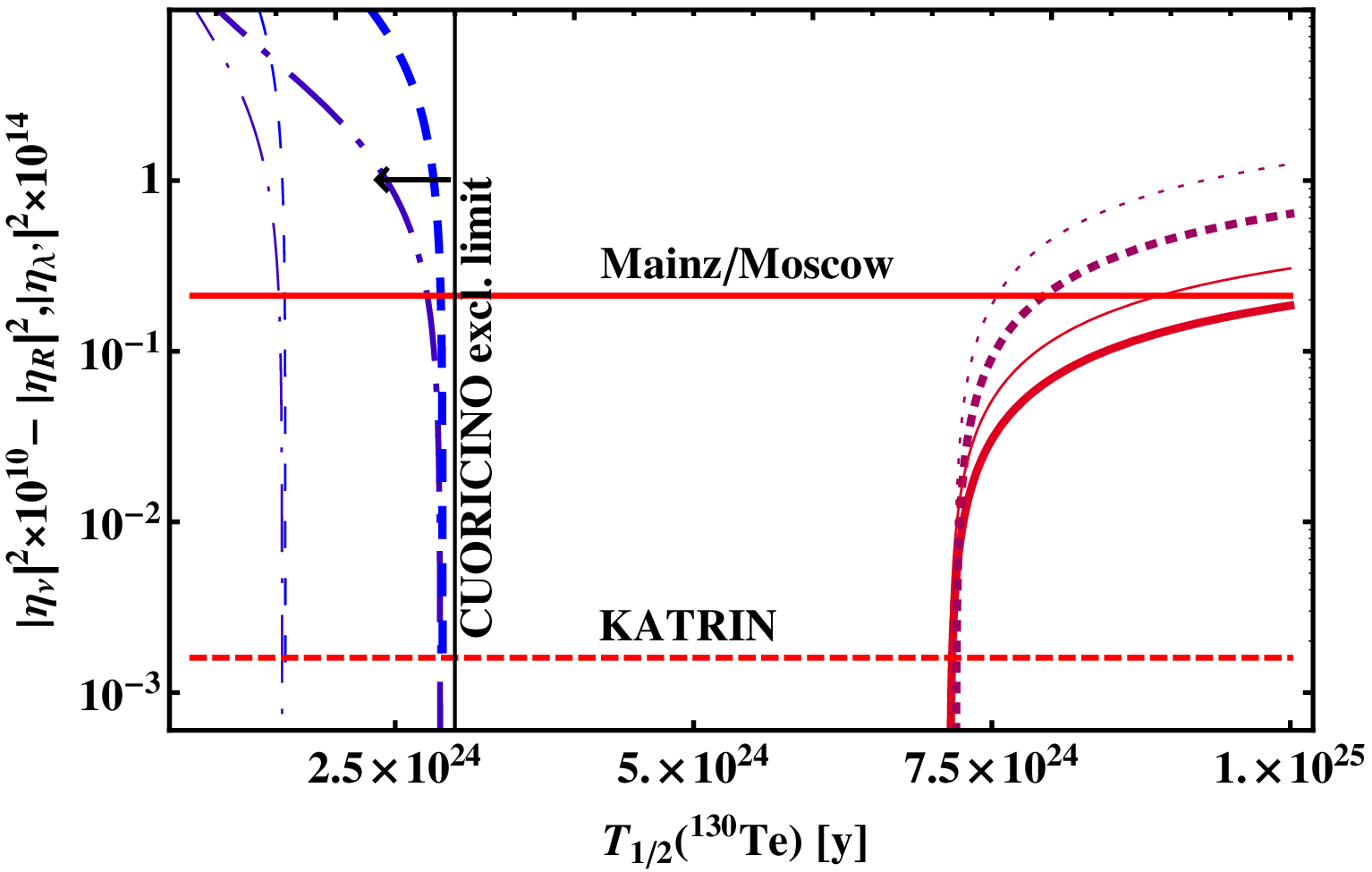}}
 \vspace{5mm}
 \subfigure
   {\includegraphics[width=7.5cm]{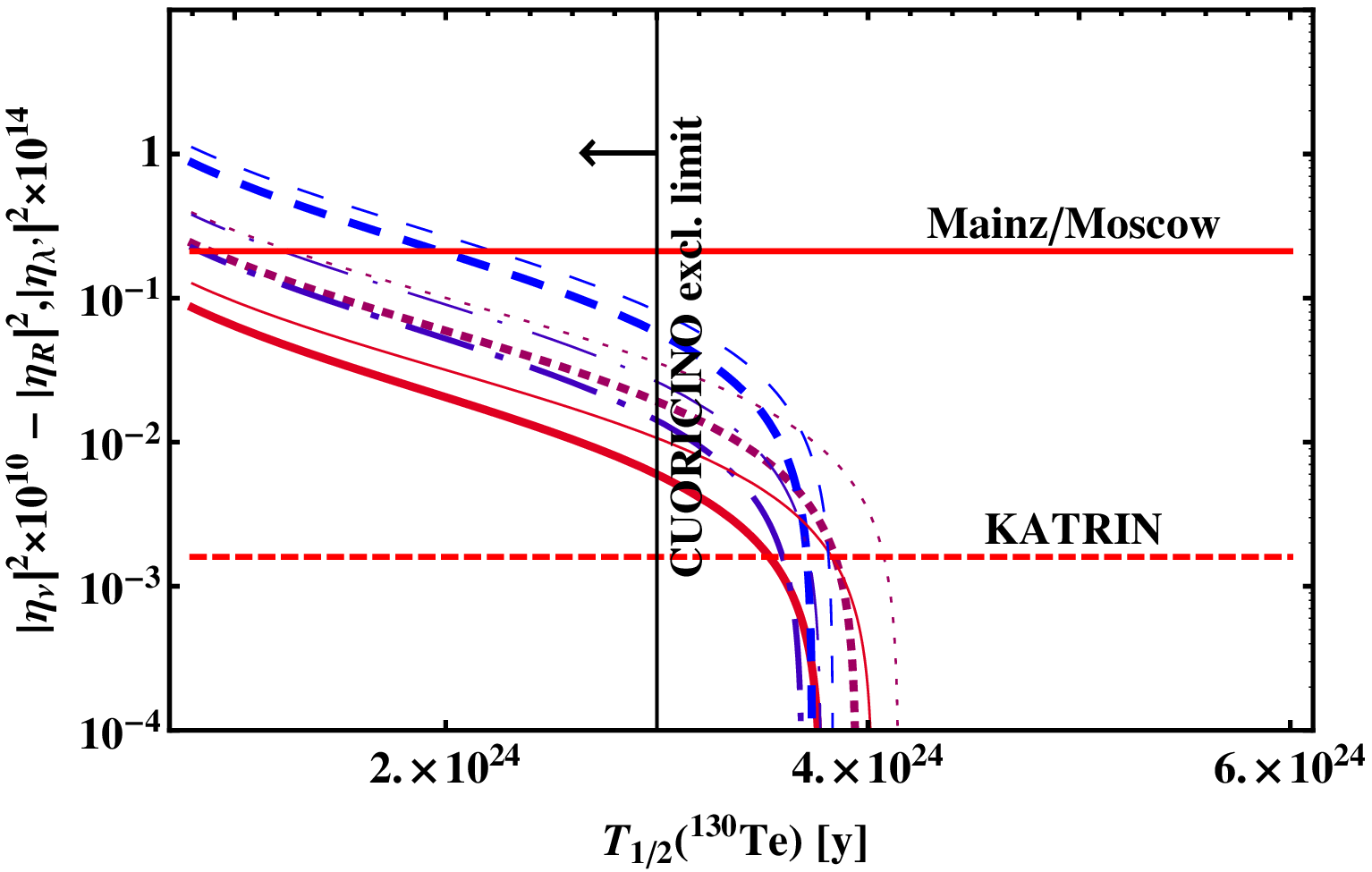}}
     \end{center}
\vspace{-1.0cm}
    \caption{\label{fig:figd1}
The parameters $|\eta_\nu|^2\times 10^{10}$ (solid line) and
$|\eta_L|^2\times 10^{14}$ (dotted line) of the light and heavy
LH Majorana neutrino exchange mechanisms, and
$|\eta_{\lambda'}|^2\times 10^{14}$ (dashed-dotted line)
and $|\eta_L|^2\times 10^{14}$ (dashed line)
of the gluino and  heavy LH Majorana
neutrino exchange mechanisms,
obtained from eq. (\ref{intsol1})
using the Argonne NMEs (left panel) and
CD-Bonn NMEs (right panel), corresponding to
$g_A=1.25$ (thick lines) and $g_A=1$ (thin lines),
for  $T^{0\nu}_{1/2}(^{76}Ge)=2.23 \times 10^{25}$y,
$T^{0\nu}_{1/2}(^{136}Xe)=1.60 \times 10^{25}$y and
letting  $T^{0\nu}_{1/2}(^{130}Te)$ free. 
See text for details.}
\end{figure}

 Next, we would like to illustrate the possibility to distinguish
between two pairs of interfering mechanisms i) A+B and ii) B+C,
which share one mechanism, namely B,
from the data on the half-lives of three isotopes.
In this case we can set two systems of three equations,
each one in three unknowns. We
will denote the corresponding LNV parameters as i) $|\eta_A|^2$,
$|\eta_B|^2$ and ii) $|\eta_B|^2$ and $|\eta_C|^2$, while the
interference parameters will be denoted as i) $z$ and ii) $z'$.
Fixing two of the three half-lives, say $T_i$ and $T_j$,
the possibility to discriminate  between the
mechanisms $A$ and $C$ relies on the
dependence of $|\eta_A|^2$ and $|\eta_C|^2$
on the third half-life, $T_k$.
Given $T_i$ and $T_j$,  it will be possible to discriminate
between the mechanisms $A$ and $C$
if the two intervals of values of $T_k$
where  $|\eta_A|^2 > 0$ and
$|\eta_C|^2 > 0$, do not overlap.
If, instead, the two
intervals partially overlap,
complete discrimination would be impossible, but
there would be a large interval of values of $T_k$
(or equivalently, positive solutions values of the
LNV parameters) that can be excluded using present or future
experimental data. In order to have
non-overlapping positive solution intervals of $T_K$,
corresponding to $|\eta_A|^2 > 0$ and $|\eta_C|^2 > 0$,
the following inequality must hold:
\be
\frac{({M'}^{0\nu}_{k,A} {M'}^{0\nu}_{i,B}-{M'}^{0\nu}_{i,A}
{M'}^{0\nu}_{k,B}) ( {M'}^{0\nu}_{k,A}
{M'}^{0\nu}_{j,B}-{M'}^{0\nu}_{j,A} {M'}^{0\nu}_{k,B})}
{({M'}^{0\nu}_{k,B} {M'}^{0\nu}_{i,C}-{M'}^{0\nu}_{i,B}
{M'}^{0\nu}_{k,C}) ( {M'}^{0\nu}_{k,B}
{M'}^{0\nu}_{j,C}-{M'}^{0\nu}_{j,B} {M'}^{0\nu}_{k,C}) }<0.
\label{conddiscr}
\ee
%
The above condition can be satisfied
only for certain sets of isotopes.
Obviously, whether it is fulfilled or not
depends on the values of
the relevant NMEs.
We will illustrate
this on the example of an
oversimplified analysis involving
the light Majorana neutrino exchange,
the heavy LH Majorana neutrino
exchange and the gluino exchange as mechanisms $A$, $B$ and $C$,
respectively, and the half-lives of $^{76}Ge$,
$^{130}Te$ and  $^{136}Xe$:  $T_1 \equiv T(^{76}Ge)$,
$T_2\equiv T(^{130}Te)$ and  $T_3\equiv T(^{136}Xe)$.
Fixing  $T_1=  2.23 \times 10^{25}$ y and
$T_3= 1.6 \times 10^{25}$ y (the EXO 90\% C.L. lower limit),
we obtain the results shown  in Fig. \ref{fig:figd1}.
As it follows from Fig. \ref{fig:figd1},
in the case of the Argonne NMEs (left panel),
it is possible to discriminate between
the standard light neutrino exchange
and the gluino exchange mechanisms:
the intervals of values of $T_2$, where the
positive solutions for the LNV parameters
of the two pairs of interfering mechanisms
considered occur, do not overlap. Further,
the physical solutions for the two LNV parameters
related to the gluino mechanism
are excluded by the CUORICINO limit on $T(^{130}Te)$ \cite{CUORI}.
This result does not change with the
increasing of $T_3$. Thus, we are lead
to conclude that for $T_3$ > 1.6 $\times$ 10$^{25}$ y
and $T_1$ given by the value claimed in
\cite{KlapdorKleingrothaus:2006ff},
of the two considered pairs of possible
interfering $\betabeta$-decay mechanisms,
only the light and heavy LH Majorana neutrino
exchanges can be generating the decay.
The solution for $|\eta_\nu|^2$ must be compatible
with the upper limit \meff < 2.3 eV \cite{MoscowH3,MainzKATRIN},
indicated with a solid horizontal line in Fig. \ref{fig:figd1}.
In the right panel of Fig. \ref{fig:figd1}
we plot also the solutions obtained with the  CD-Bonn NMEs.
In this case is not possible to discriminate between the two
considered pair of mechanisms since the condition in eq. (\ref{conddiscr})
is not satisfied.

\begin{figure}
  \begin{center}
 \subfigure
 {\includegraphics[width=7.5cm]{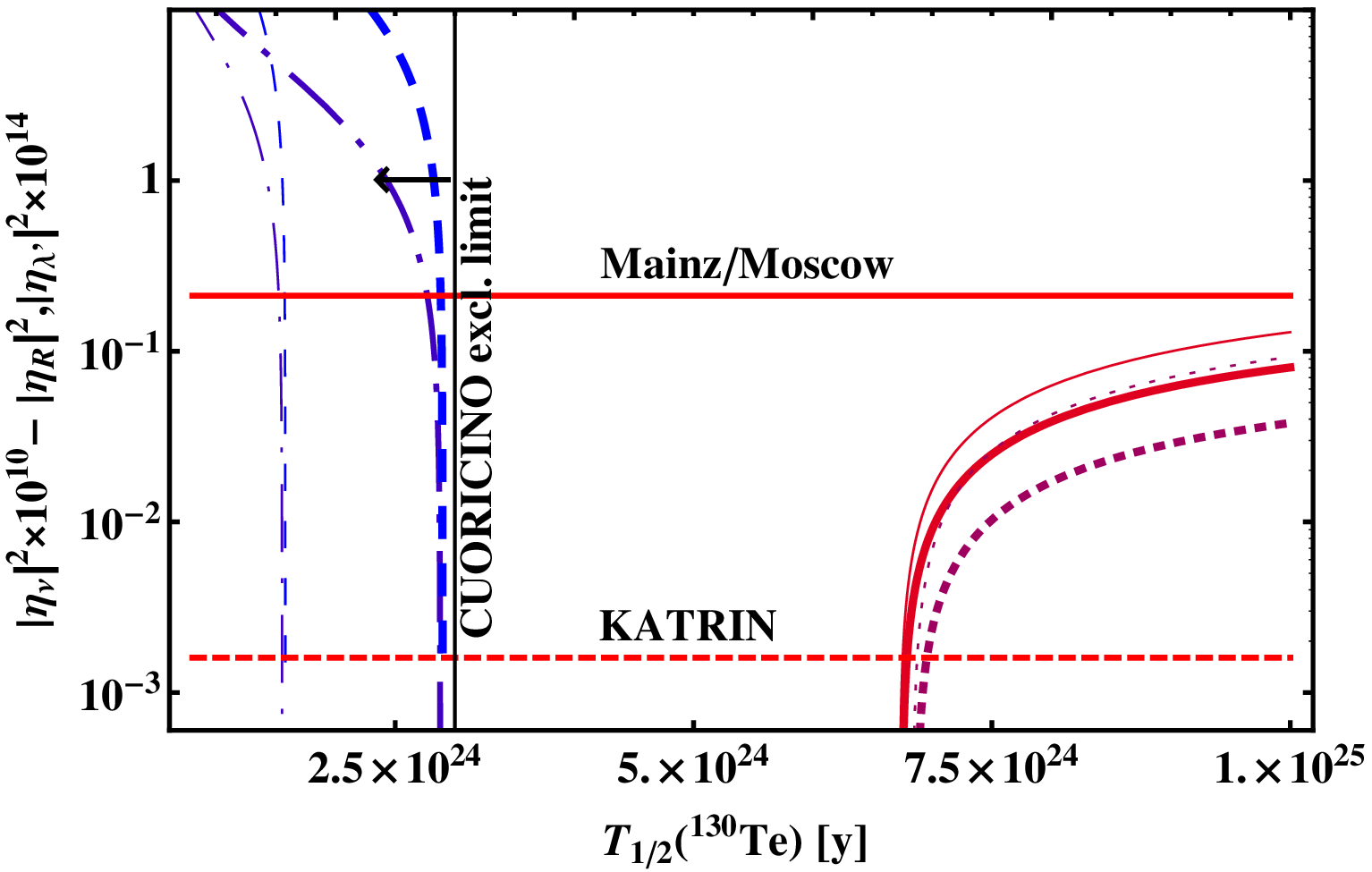}}
 \vspace{5mm}
 \subfigure
   {\includegraphics[width=7.5cm]{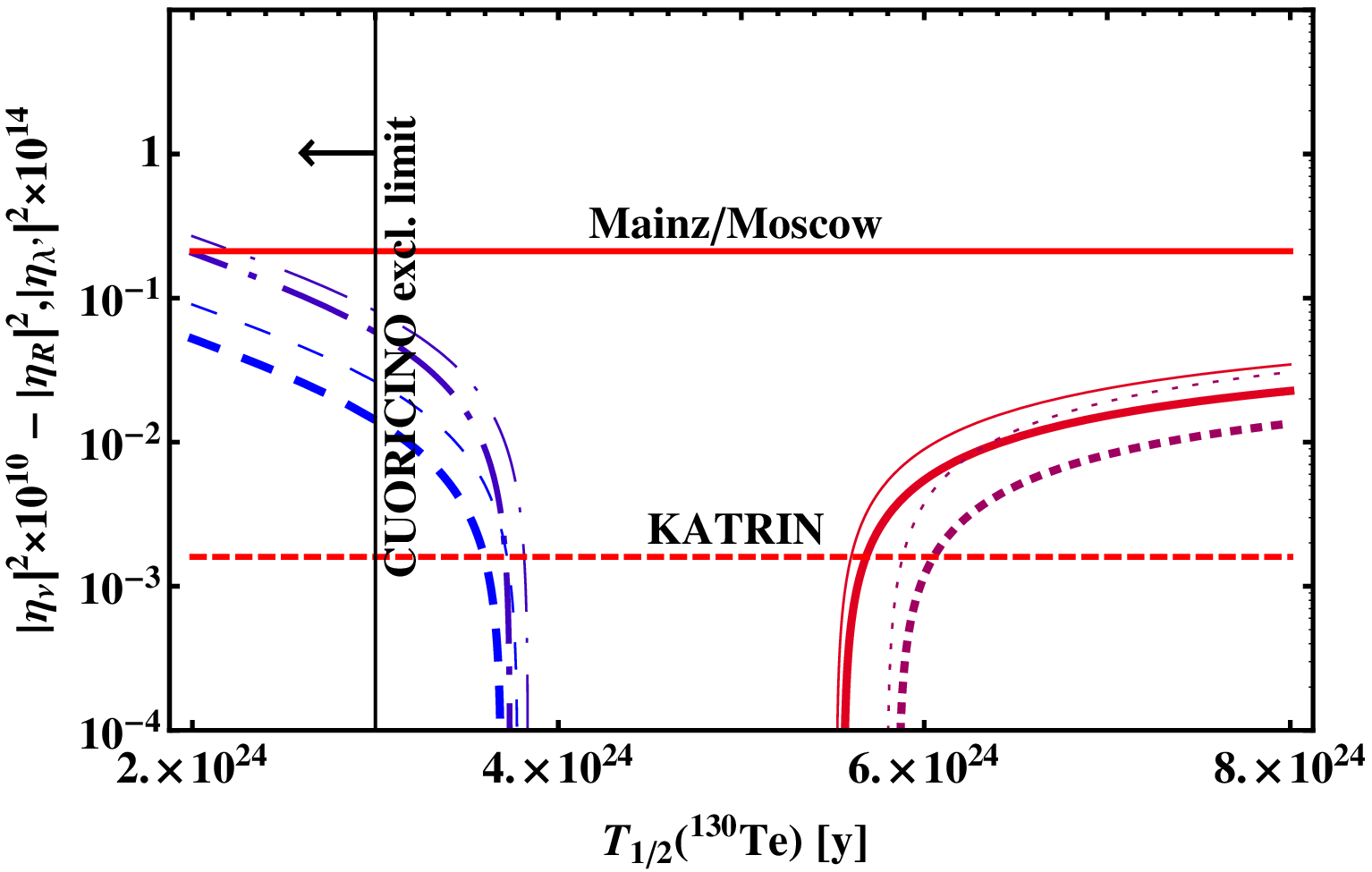}}
     \end{center}
\vspace{-1.0cm}
    \caption{\label{fig:figd2}
The same as in Fig. \ref{fig:figd1}, but for
i) $|\eta_\nu|^2\times 10^{10}$ (thick solid line) and
$|\eta_{\lambda'}|^2\times 10^{14}$ (thick dotted line) of the
 light neutrino and gluino exchange mechanisms, and
ii) $|\eta_L|^2\times 10^{14}$ (thick dashed-dotted line)
and $|\eta_{\lambda'}|^2\times 10^{14}$ (thick dashed line) of the
heavy  LH Majorana neutrino and gluino exchange mechanisms,
and using $T^{0\nu}_{1/2}(^{76}Ge)=2.23 \times 10^{25}$ y and
$T^{0\nu}_{1/2}(^{136}Xe)=1.60 \times 10^{25}$ y. 
See text for details.
}
\end{figure}

  Another interesting example is the case in which $A$ is the light
Majorana neutrino exchange, $B$ is the gluino exchange and
$C$ the heavy LH Majorana neutrino exchange, i.e., we
try to discriminate between i) the light neutrino
plus gluino exchange mechanisms, and
ii) the heavy LH Majorana neutrino plus gluino exchange mechanisms.
We fix, like in the previous case, the values for
$T_1=  2.23 \times 10^{25}$y and $T_3= 1.6\times 10^{25}$y.
The results of this analysis are plotted in
Fig. \ref{fig:figd2}. Since the condition
in eq. (\ref{conddiscr}) is now satisfied
for NMEs obtained either with the Argonne potential or
with the CD-Bonn potential,
in this case it is possible, in principle,
to discriminate between the the two pair of mechanisms independently
of the set of NMEs used (within the sets considered by us).
This result does not change with the increasing of $T_3$.
Hence, as far as $T_1$ is fixed to the value
claimed in \cite{KlapdorKleingrothaus:2006ff} and the limits in
eq. (\ref{limit}) are satisfied,
the two intervals of values of $T_2$, in which the ``positivity
conditions'' for i) $|\eta_\nu|^2$, $|\eta_{\lambda'}|^2$ and
$z$, and for ii) $|\eta_{\lambda'}|^2$, $|\eta_N|^2$ and $z'$,
are satisfied, are not overlapping  (Fig. \ref{fig:figd2}).

%
\section{Conclusions and Summary}
%

 We have investigated the possibility to discriminate between
different pairs of CP non-conserving mechanisms inducing the
neutrinoless double beta $\betabeta$-decay by using
data on $\betabeta$-decay half-lives of nuclei
with largely different nuclear matrix elements (NMEs).
The mechanisms studied are: light Majorana neutrino exchange,
heavy left-handed (LH) and heavy right-handed (RH)
Majorana neutrino exchanges, lepton charge non-conserving
couplings in SUSY theories with $R$-parity breaking giving rise
to the ``dominant gluino exchange'' and the ``squark-neutrino''
mechanisms. Each of these mechanisms is characterized
by a specific lepton number violating (LNV) 
parameter  $\eta_{\kappa}$, where the index $\kappa$ labels 
the mechanism. For the five mechanisms listed above 
we use the notations 
$\kappa = \nu,L,R,\lambda',\tilde{q}$, respectively. 
The parameter $\eta_{\kappa}$ will be complex,
in general,  if the mechanism $\kappa$
does not conserve the CP symmetry.
The nuclei considered are
$^{76}$Ge, $^{82}$Se, $^{100}$Mo, $^{130}$Te
and $^{136}$Xe. Four sets of nuclear matrix
elements (NMEs) of the $\betabeta$-decays of these
five nuclei, derived within the
Self-consistent Renormalized
Quasiparticle Random Phase Approximation
(SRQRPA), were employed in our analysis.
They correspond to two types of nucleon-nucleon 
potentials - Argonne (``Argonne NMEs'') and CD-Bonn 
(``CD-Bonn NMEs''), and two values 
of the axial coupling constant $g_A = 1.25;1.00$.
Given the NMEs and the phase space factors of the 
decays, the half-life of a given nucleus depends on 
the parameters  $|\eta_{\kappa}|^2$ of the mechanisms 
triggering the decay (eq. (\ref{solnonint})). 

  In the present article we have considered in detail 
the cases of two non-interfering and two interfering 
mechanisms inducing the $\betabeta$-decay.
If two non-interfering mechanisms $A$ and 
$B$ cause the decay, the parameters 
$|\eta_{A}|^2$ and $|\eta_{B}|^2$
can be determined from data on the half-lives 
of two isotopes, $T_1$ and $T_2$ as 
solutions of a system of two linear equations. 
If the half-life of one isotope is known, 
say $T_1$, the positivity condition which the solutions 
$|\eta_{A}|^2$ and $|\eta_{B}|^2$ must satisfy,  
$|\eta_{A}|^2 \ge 0$ and $|\eta_{B}|^2 \ge 0$, 
constrain the half-life of the second isotope $T_2$ 
(and the half-life of any other isotope for that matter) 
to lie in a specific interval \cite{FMPSV0311}.
If $A$ and $B$ are interfering mechanisms, 
$|\eta_{A}|^2$ and $|\eta_{B}|^2$ and the
interference term parameter, 
$z _{AB} \equiv 2\cos\alpha_{AB}|\eta_{A}\eta_{B}|$
which involves the cosine of an unknown relative 
phase $\alpha_{AB}$ of $\eta_{A}$ and $\eta_{B}$, 
can be uniquely determined, in principle, 
from data on the half-lives of three nuclei, $T_{1,2,3}$.
In this case, given the half-life of one isotope, say $T_1$, 
the ``positivity conditions'' 
$|\eta_{A}|^2 \ge 0$, $|\eta_{B}|^2 \ge 0$ 
and $-1\leq \cos\alpha_{AB} \leq 1$ constrain 
the half-life of a second isotope, say $T_2$, to 
lie in a specific interval, and the half-life 
of a third one, $T_3$, to lie in an interval which 
is determined by the value of $T_1$ and the 
interval of allowed values of $T_2$.

  For all possible pairs of non-interfering mechanisms 
we have considered (light, or heavy LH Majorana neutrino, 
and heavy RH Majorana neutrino exchanges;  
gluino, or squark-neutrino, and RH Majorana neutrino 
mechanisms), these ``positivity condition'' intervals
of values of $T_2$ were shown in \cite{FMPSV0311} 
to be essentially degenerate if $T_1$ and $T_2$ 
correspond to the half-lives of any pair of the 
four nuclei $^{76}$Ge, $^{82}$Se, $^{100}$Mo and $^{130}$Te.
This is a consequence of the fact that 
for each of the five single mechanisms discussed, 
the NMEs for $^{76}$Ge, $^{82}$Se, $^{100}$Mo and 
$^{130}$Te differ relatively little 
\cite{FMPSV0311,ELisietalMM11}: 
the relative difference between the NMEs of any two 
nuclei does not exceed 10\%. One has similar degeneracy 
of ``positivity condition'' intervals $T_2$ and $T_3$
in the cases of two constructively interfering mechanisms 
(within the set considered). These degeneracies might 
irreparably plague the interpretation of the $\betabeta$-decay 
data if the process will be observed.

  The NMEs for $^{136}Xe$,  results of calculations of which 
using the SRQRPA method are presented in the present article, 
differ significantly from those of $^{76}$Ge, $^{82}Se$, 
$^{100}$Mo and $^{130}$Te, being by a 
factor $\sim (1.3 - 2.5)$ smaller. As we have shown in 
the present article, this allows to lift to a certain 
degree the indicated degeneracies and to draw conclusions 
about the pair of non-interfering (interfering) 
mechanisms possibly inducing the $\betabeta$-decay
from data on the half-lives of $^{136}Xe$ and
of at least one (two) more isotope(s) which can
be, e.g., any of the four, $^{76}Ge$, $^{82}Se$, 
$^{100}Mo$ and $^{130}Te$ considered.

 We have analyzed also the possibility to discriminate 
between two pairs of non-interfering (or interfering) 
$\betabeta$-decay mechanisms when the pairs have 
one mechanism in common, i.e., 
between the mechanisms i) $A$ + $B$ and ii) $C$ + $B$, 
using the half-lives of the same two isotopes. We have 
derived the general conditions under which it would 
be possible, in principle, to identify which pair 
of mechanisms is inducing the decay (if any).
We have shown that the conditions of interest 
are fulfilled, e.g., 
for the following two pairs of non-interfering mechanisms  
i) light neutrino exchange (A) and heavy RH Majorana 
neutrino exchange (B) and ii) gluino exchange (C) and 
heavy RH Majorana neutrino exchange (B), 
and for the following two pairs of interfering mechanisms 
i) light neutrino exchange (A) and heavy 
LH Majorana neutrino exchange (B) and 
ii) gluino exchange (C) and heavy 
LH Majorana neutrino exchange (B), 
if one uses the Argonne NMEs in the analysis.
They are fulfilled for both the Argonne NMEs and CD-Bonn NMEs, 
e.g., for the following two pairs of interfering mechanisms
i) light neutrino exchange (A) and gluino exchange (B), and 
ii) heavy LH Majorana neutrino exchange (C) and  gluino exchange (B).

  We have also exploited the implications of the EXO lower
bound on the half-life of $^{136}Xe$ for the problem 
studied. We have shown, in particular, that for all four pairs 
of non-interfering mechanisms considered and the Argonne 
NMEs, the half-life of $^{76}Ge$ claimed in 
\cite{KlapdorKleingrothaus:2006ff} is incompatible with 
the EXO lower bound on the half-life of  $^{136}Xe$ 
\cite{Auger:2012ar}.
If we use the CD-Bonn NMEs instead, we find that the result 
half-life of $^{76}Ge$ claimed in \cite{KlapdorKleingrothaus:2006ff} 
is compatible with the EXO lower bound on 
the half-life of  $^{136}Xe$ for values of the 
corresponding LNV parameters lying in extremely 
narrow intervals.

  To summarize, the results obtained in the 
present article show that using the $\betabeta$-decay 
half-lives of nuclei with largely different 
nuclear matrix elements would help resolving the problem 
of identifying the mechanisms triggering the decay.

\section*{Acknowledgments}

This work was supported in part by the INFN program on
``Astroparticle Physics'', the Italian MIUR program on
``Neutrinos, Dark Matter and  Dark Energy in the Era of LHC'', 
the World Premier International Research Center 
Initiative (WPI Initiative), MEXT, Japan  (S.T.P.) and 
by the European Union FP7-ITN INVISIBLES 
(Marie Curie Action, PITAN-GA-2011-289442).
 F. \v S acknowledges the support by the VEGA Grant agency 
of the Slovak Republic under the contract No. 1/0876/12.


\end{document}